\documentclass[epsfig]{article}

\usepackage{makeidx}
\usepackage{graphicx}

\newcommand{\refcompcomp}{2.1}
\newcommand{\refpretlus}{2.3}
\newcommand{\refnew}{4}
\newcommand{\refnewat}{4.2}
\newcommand{\refnewion}{4.3}
\newcommand{\refstrateg}{5}
\newcommand{\refexamphhe}{6.2}

\newcommand{\refexampbstar}{6.3}
\newcommand{\refexampoptab}{6.4}
\newcommand{\refexampcwd}{6.5}
\newcommand{\refexampdisk}{6.6}
\newcommand{\refnonst}{7}
\newcommand{\refnonstglob}{7.4.1}
\newcommand{\refnstexaoptab}{7.10.3}
\newcommand{\refmodin}{8}
\newcommand{\refinpchan}{8.2}
\newcommand{\refout}{9}
\newcommand{\reftrouble}{10}
\newcommand{\refions}{11}
\newcommand{\refionsup}{11.4}
\newcommand{\refnsttwophys}{12}
\newcommand{\refnsttwonum}{13}
\newcommand{\refnsttwooptab}{13.1}
\newcommand{\reflist}{16}
\newcommand{\refnsttwohyd}{12.1.1}

\newcommand{\refnsttwoopadd}{12.2}
\newcommand{\refnsttwoeos}{12.7}

\newcommand{\refaccel}{3.4}

\makeindex

\begin{document}

\title{\bf A brief introductory guide to {\sc tlusty} and {\sc synspec}}  
\author{I. Hubeny\footnote{University of Arizona, 
Tucson; USA; hubeny{\tt @}as.arizona.edu}~ and 
T. Lanz\footnote{Observatoire de C\^{o}te d'Azur, France; {\tt lanz{\tt @}oca.eu}}}
\date{\today}
\maketitle

\begin{abstract}
This is the first of three papers that present a detailed guide for working with
the codes {\sc tlusty} and {\sc synspec} to generate model stellar atmospheres
or accretion disks, and to produce detailed synthetic spectra. 
In this paper, we present a very brief manual intended for casual users who intend
to use these codes for simple, well defined tasks. This paper does not present
any background theory, or a description of the adopted numerical approaches,
but instead uses simple examples to explain how to employ these codes. In particular,
it shows how to produce a simple model atmosphere from the scratch, or how to
improve an existing model by considering more extended model atoms. This
paper also presents a brief guide to the spectrum synthesis program {\sc synspec}.
\end{abstract}

\tableofcontents

\newpage

\section{Introduction}
\label{intro}

Unlike previous manuals, this one is much more comprehensive and is consequently
divided into three parts.
This paper, the first part, presents a very brief manual for busy or impatient users.
It shows how to work with {\sc tlusty} and {\sc synspec} on simple practical examples.  
In particular, it shows how to use {\sc tlusty} to produce simple model 
atmospheres from the scratch, as well as to modify existing models from
available grids to change basic parameters, or to extend model atoms to
more complex ones. It also presents a brief guide to the spectrum synthesis 
program {\sc synspec} and associated programs to obtain detailed predicted
spectra to be compared to observations.

The second part (Hubeny \& Lanz 2017b; hereafter called Paper II), 
presents a detailed reference manual for {\sc tlusty}, which contains
a description of basic physical assumptions and equations used to
model an atmosphere, together with an overview of the numerical
methods to solve these equations.

The third part (Hubeny \& Lanz 2017c; hereafter called Paper III) represents
a basic operational manual for {\sc tlusty}. It
provides a guide for understanding the essential 
features and the basic modes of operation of the program.
Since {\sc tlusty} offers great many numerical options, as well as various
approximations of the basic physical description, there are also many options
the code can consider, and consequently many switches and parameters
which they control, Paper III is divided into two parts. The first part describes the most
important input parameters and available numerical options.
The second part
covers additional details and a comprehensive description
of all physical and numerical options, and a description of all input parameters, 
many of which needed only in special cases.

The user who intends to construct simple models, or perhaps more sophisticated
models using samples of input data provided in the {\sc tlusty} website by just 
modifying them, will obtain
enough information and guidance from this Paper~I. However, in case of problems,
for instance a slow convergence of the iteration process, or the lack
thereof, the user should consult appropriate sections of other parts.
Dedicated users are encouraged to consult the full content of all three papers.

Both programs, {\sc tlusty} and {\sc synspec}, together with the utility programs
{\sc pretlus} and {\sc rotin}, and with all the input files needed to run the test case
models, are available either on the main {\sc tlusty} website\footnote{http://tlusty.oca.eu/},
or as a gzipped tar file that contains the whole 
distribution\footnote{http://aegis.as.arizona.edu/$\,\,\widetilde{\ }\,$hubeny/pub/tlusty205.tar.gz}.

\section{Basic characteristics}
\label{intro_gen}

The present set of computer programs is a package 
designed to accomplish a wide range of stellar spectroscopic diagnostics studies. 
It can construct a model atmosphere either from scratch or from another model,
compute a detailed spectrum on a selected wavelength range, and, if desired,
convolve the synthetic spectrum with an arbitrary rotational velocity and instrumental
profile of the spectrograph to obtain a predicted spectrum directly comparable to
observations. 

\subsection{{\sc tlusty}}
The basic component of the package is {\sc tlusty}, a program for calculating
plane-parallel, horizontally homogeneous model stellar atmospheres in
hydrostatic  and radiative (or radiative + convective) equilibrium. 
Departures from local thermodynamic 
equilibrium (LTE) are allowed for a set of occupation numbers (populations) 
of selected atomic and ionic energy levels.
Thanks to the flexibility and efficiency of the adopted numerical procedures,
described in detail in Paper~II, the number of such levels, referred to as  
{\em NLTE levels} can be essentially as large as the amount of available atomic data
allows.

The code offers several global numerical schemes to solve the set of structural equations;
most importantly the hybrid complete-linearization/ accelerated Lambda-iteration
(CL/ALI) method, or the Rybicki reorganization scheme, both augmented by
applying various mathematical acceleration procedures.  The program
employs the concept of superlevels and superlines, thus allowing for
computing fully consistent, non-LTE metal line-blanketed model atmospheres.
For cool objects for which LTE models are satisfactory, the code
can use pre-calculated opacity tables that include all important
atomic and molecular opacity sources.

The program is fully data oriented as far as the choice of
atomic species, ions, energy levels, transitions, and opacity sources
is concerned.  In several cases, there are various default formulas and internal
tables provided for selected cross sections and opacity sources, but in most
cases the user is able to choose between several options. On top of that,
there is a collection of input atomic data for a number of atoms and ions
accessible from the {\sc tlusty} website,, 
so users are not obliged to collect the necessary atomic data themselves.

In the past, there was a separate variant called {\sc tlusdisk},
designed to calculate vertical structure of accretion disks. Starting
at version 200, both programs are contained within a single,
universal code {\sc tlusty}. The user is able to make a choice between
computing a model of a stellar atmosphere or a vertical structure of
an accretion disk, based on input data. In the following text, we will usually
refer to a ``stellar atmosphere'', but everything applies, unless
specifically noted, for an accretion disk as well.

The code is written in standard FORTRAN77, making it highly
portable. The current version removed outdated features of FORTRAN, and 
adheres  to modern standards, so the program compiles on all platforms, most
importantly using compilers available within the LINUX and Mac OSX operation
systems.

\subsection{{\sc synspec}}
\label{intro_ass}

The second basic program is {\sc synspec}, current version 51,
\index{SYNSPEC program}
which is a program for calculating the spectrum emergent from a given model
atmosphere. It was originally described in Hubeny, Lanz, \& Jeffery (1994),
also see Hubeny \& Lanz (2011), and a detailed description and the user's 
manual is available on the {\sc tlusty} website.
It has been originally designed to synthesize emergent spectra from model atmospheres
calculated using {\sc tlusty}, but may also be used with other model atmospheres
as input, for instance Kurucz's {\sc atlas} models. The program is complemented by the
program {\sc rotin} which calculates the rotational and instrumental
convolutions for the net spectrum produced by {\sc synspec}.

\subsection{Associated programs}
\label{intro_ass}

Finally, there is a number of interface and utility programs. They provide 
a graphical
interface for plotting output models, the convergence log etc. (written in
IDL). Similarly, there is graphical interface {\sc synplot}  (written
\index{SYNPLOT program}
again in IDL), which enables the user to perform
an interactive work with {\sc synspec} 
and its utility programs (computing the spectrum, plotting it, performing
various convolutions, and identifying and annotating predicted lines). It is available 
online\footnote{http://aegis.as.arizona.edu/$\,\,\widetilde{\ }\,$hubeny/pub/synplot2.1.tar.gz},
together with a detailed user's manual.

There is also a program {\sc pretlus}, distributed along the main {\sc tlusty}
\index{PRETLUS program}
code, which can be run before running {\sc tlusty}, and which prints
dimensions of the most important arrays. Its results can be used to
set up the array dimensions in {\sc tlusty}, and thus to control the overall
memory consumption of {\sc tlusty}. Its operation and function is
described in detail in Paper III, \S\,\refpretlus.

\subsection{Why use {\sc tlusty}? }
\label{intro_why}

We believe that the code is useful  because it was 
designed to be very flexible, versatile, but also easy to use. 
Such a goal presents a non-trivial task,
because with increasing flexibility and versatility of a program, the complexity
of operation and the difficulty of its usage usually increases as well.

{\sc tlusty} can construct models ranging from the simplest ones, being computed
just in seconds, and thus having a potential to be used as an interactive 
pedagogical tool, to the most complex NLTE metal line-blanketed models
whose construction may easily take hours to tens of hours on modern
computers.

{\sc tlusty} is also extremely versatile. It can construct models for stellar 
atmospheres, or a vertical structure of accretion disks.
As far as the range of physical conditions is concerned,
{\sc tlusty} can compute model atmospheres
from extremely cool brown dwarfs or giant planets, to very hot 
accretion disks around solar-mass black holes. In other words, one can
model structures with temperatures from a few times of $10^1$ K
to a few times of $10^8$ K. The limit on the low-temperature side is not
imposed by algorithmic or physical shortcomings, but rather by the availability
of the corresponding available molecular opacity data. On the high-temperature
side, the limiting factor is the current lack of fully relativistic treatment of the 
Compton scattering, and a more accurate description of inner-shell transitions,
pair production, and other high-energy phenomena..
For densities, there is no strict limit on a low-density side; the limit on the
high-density side is the total particle number density of about $10^{20}$~cm${}^{-3}$,
above which the multi-particle effects in the equation of state, atomic transition rates, 
and description of line broadening, became important.
 
What are the basic limitations of {\sc tlusty}? Besides the limitations of
temperature and density, which usually are not very serious for most applications,
the most stringent limitations follow from the basic assumptions and
approximations.
It is a 1-D code that assumes a plane-parallel, horizontally-homogeneous 
atmosphere. It is therefore inapplicable for highly inhomogeneous media,
although it can still be used to describe atmospheres with large patches
with different physical conditions, provided that those patches have large 
enough horizontal optical
depths that they are independent of the ambient atmosphere. {\sc tlusty} is not
applicable for spherical atmospheres, although its adaptation to spherical 1-D
geometry would be relatively straightforward. (Actually, spherical version
existed in the past, but did not propagate to later versions).

The second basic limitation is the assumption of hydrostatic equilibrium 
(or vertical hydrostatic equilibrium in the case of disks). {\sc tlusty} is therefore
inapplicable for instance for stellar winds, although it can still model
deeper layers of a star that exhibits a wind in its outer layers. The hydrostatic
models produced by {\sc tlusty} would still be useful for the atmospheric layers
where the outflow velocity does not significantly exceed the thermal velocity.

The third basic limitation is the assumption that the only external force
acting on the material is gravity. 
Therefore, {\sc tlusty} is inapplicable, or its results should be viewed
with caution, for instance for atmospheres with strong magnetic fields,
or for atmospheres of rapidly rotating stars where the centrifugal force
plays an important role.

\subsection{Word of caution}
\label{intro_word}

Although considerable effort has been devoted to eliminate errors in the
code, there is by no means a guarantee that it is error free. 
In fact, the same warning as Bob Kurucz usually puts into his codes applies
here as well: ``This code is guaranteed to contain errors.''
The user is thus warned against using the program as a ``black box".
It would be highly appreciated if any errors detected by the user, and
any comments or suggestions for improvements, are communicated to
I. Hubeny ({\tt hubeny@as.arizona.edu}) or T. Lanz
({\tt lanz@oca.eu}).


\section{Installing the package}

As mentioned above, the program together with the  necessary data files and 
the input files for several
examples and test cases can be obtained either from \\ [2pt]  
$\bullet$ the main {\sc tlusty} site
in Observatoire de Cote d'Azur in Nice, 

{\tt http://tlusty.oca.eu}, \\ [2pt]
which is a website previously located at {\tt http://nova.astro.umd.edu}, or \\ [2pt]
$\bullet$  the tarball of the whole package obtained from 

{\tt http://aegis.as.arizona.edu/$\,\widetilde{\ }\,$hubeny/pub/tlusty205.tar.gz}.\\ [2pt]
In the following text, this site will be referred to as the {\em Arizona site}.
When using the Nice site, one can download the programs, data, and examples
individually, while by downloading and extracting the tarball obtained from the
Arizona site one obtains all the files, together with  a specific directory tree where
the files are located.

The main website at Nice is, at the time of submitting this manuscript (June 2017),
not yet fully operational. When it is, the website will contain detailed instruction 
for downloading. In the meantime, or as an alternative means, 
the potential user is encouraged to download the package from the Arizona site.
One can either download the file(s) interactively, or using the LINUX command 
{\tt wget}, if available. In the subsequent text, we will use the {\tt wget}
command as an example, keeping in mind that the files can also be downloaded
interactively using any web browser.

Complete download proceeds in one or more steps, depending on the user's
needs. The first file is the main one, required to be downloaded in any case.
In addition, there are several auxiliary files which are optional, The reasons for having several
files to download are several: (i) The auxiliary files are big, which may cause problems in downloading all the files together; (ii) long-time users may already have some or all of the
auxiliary files; and (iii) many applications do not require these files, so for some 
users they would represent an unnecessary waste of memory and time for downloading.
We summarize the files to download below.

\subsection{Main file}
The main file, which is sufficient for most purposes, is downloaded as follows
(we stress again that it can be downloaded interactively using any web browser):
\begin{verbatim}
wget http://aegis.as.arizona.edu/~hubeny/pub/tlusty205.tar.gz
gunzip tlusty205_tar.gz
tar -xvf tlusty205.tar
\end{verbatim} 
That action creates directory {\tt ./tlusty205}, with several subdirectories, which we 
will call here the ``standard {\sc tlusty} directory".
Although the user can obviously use any location of the files, we will use in the
following examples the directory tree that originates from the {\tt tlusty205.tar} file.
This directory also contains several shell scripts that work
with the directory tree thus generated. In the following text,.we will call this
directory tree as the {\em standard directory tree}.
If the user prefers, or has already created
a different directory system, the provided script files have to be modified accordingly. 
In any case, the test cases can be run either individually, or all of them in one run. 

The standard directory tree has the following structure: The main directory 
{\tt ./tlusty205} contains the shell scripts {\tt RTlusty} and {\tt RSynspec} for 
running {\sc tlusty}, and {\sc synspec}, respectively, and five subdirectories:
\begin{itemize}
\item {\tt tlusty} - which contains the source file for the current version of {\sc tlusty},
called {\tt tlusty205.f}, together with auxiliary files {\tt *.FOR} - see below;
\item {\tt synspec} - an analogous directory for {\sc synspec}. It contains the
source files, plus the utility program {\sc rotin};
\item {\tt pretlus} - source files for the utility program {\sc pretlus} -see Paper~III,
\S\,\refpretlus;
\item {\tt data} - a collection of needed atomic data.
\item {\tt examples} - it contains the shell script {\tt Runtest} that runs all the test cases, and five
subdirectories for the test cases, described in detail in Paper~III, Chap. 6.
\begin{itemize}
\item {\tt hhe} -- a simple H-He LTE and NLTE model atmosphere - see \S\,\ref{testc} here 
and Paper~III, \S\,\refexamphhe.
\item {\tt bstar} -- a NLTE metal line-blanketed model of a B stars, analogous to models
from the {\sc bstar2006} grid (Lanz \& Hubeny 2007) -- see Paper~III, \S\,\refexampbstar.
\item {\tt optab} -- an LTE line-blanketed model of a K stars, using an opacity table - see
Paper~III, \S\,\refexampoptab.
\item {\tt cwd} -- an LTE and NLTE model of a moderately cool DA (pure-hydrogen)
white dwarf - see Paper~III, \S\,\refexampcwd.
\item {\tt disk} - a simple H-He LTE and NLTE model of a vertical structure if an
accretion disk - see Paper~III, \S\,\refexampdisk.
\end{itemize}
All subdirectories contain (i) input data for a run, and (ii) some output files obtained
by running the test cases on a MacBook Pro done by the author (I.H.), which are 
included for comparisons with user's runs to the purposes of checking an implementation
of the codes on the user's platform.
\end{itemize}

\subsection{Opacity table}
The users who intend to compute LTE metal line blanketed model atmospheres,
using a pre-calculated opacity table whose application is shown in Paper~III, 
\S\,\refexampoptab, 
need to download the file {\tt absopac.dat} into
directory {\tt data}. One proceeds as follows:
\begin{verbatim}
cd ./tlusty205/data
wget http://aegis.as.arizona.edu/~hubeny/pub/absopac.dat.gz .
gunzip absopac.dat.gz
\end{verbatim}

\subsection{Iron data}
If the user intends to calculate NLTE metal line blanketed models for early
type stars, as shown on the test case presented in Paper~III, \S\,\refexampbstar,
and using the input data stored  in \${\tt TLUSTY/examples/bstar}, one needs
to download detailed data for iron. One proceeds as follows:
\begin{verbatim}
cd ./tlusty205/data
wget http://aegis.as.arizona.edu/~hubeny/pub/gf26.tar.gz .
gunzip gf26.tar.gz
tar -xvf gf26.tar
\end{verbatim}
As a results, the files {\tt gf2601.gam} to {\tt gf2605.gam}, and 
analogously for and {\tt gf26*.lin} are created that
represent Kurucz data files for iron levels and transitions, and from which
{\sc tlusty} generates necessary parameters for iron superlevels and 
superlines - see Paper~II, \S\,3.6.

\subsection{Line lists}
Finally, for computing detailed synthetic spectra with {\sc synspec}, one
needs full line list(s).  We stress that the standard "{\tt data}" directory
contains a sample line list for tests, with line data for wavelengths
between 1390 and 1500 \AA. When computing synthetic spectra for
different spectral regions, one needs to obtain one of more of the 
following files.
They should also be placed in the "data" directory, namely
The user can choose from several possibilities:\\ [-2 pt]

$\bullet$ full line list between 1 \AA~ and about 15 microns.
The list contains about $5.5\times 10^6$ lines (download  is about 121 MB).
\begin{verbatim}
cd ./tlusty205/data
wget http://aegis.as.arizona.edu/~hubeny/pub/gfTOT.dat.gz .
\end{verbatim}

$\bullet$ lines in the FUV region, between 880 and 2,000 \AA,
The list contains about 121,000 strongest lines in this region.
\begin{verbatim}
cd ./tlusty205/data
wget http://aegis.as.arizona.edu/~hubeny/pub/gfFUV.dat.gz.
\end{verbatim}

$\bullet$ lines only in the visible region, between 3,000 and 7,500 \AA,
The list contains about 149,.000 strongest lines in this region.
\begin{verbatim}
cd ./tlusty205/data
wget http://aegis.as.arizona.edu/~hubeny/pub/gfVIS99.dat.gz.
\end{verbatim}

Other line lists (for instance the infrared region, or molecular lines) are in 
principle available, but are not a part of the standard distribution.


\section{Running {\sc tlusty}}

\subsection{Source code files and compilation}

The program is distributed as one big file that contains all subroutines,
plus several (8) small files that contain the {\tt INCLUDE} statements which
declare most of variables and arrays, together with array dimensions.
These can be used to recompile the code to decrease or increase its
memory consumption, as explained in Paper III, \S\,\refcompcomp.  
All files should reside in the same directory.

A compilation of the program is explained in more detail on Paper III,
\S\,\refcompcomp, where the compiler instructions on different
platforms are summarized. Here we only show a compilation using
{\tt gfortran}, which is available on most Mac and LINUX platforms,
\begin{verbatim}
        gfortran -fno-automatic [-O3] -o tlusty.exe tlusty205.f   
\end{verbatim}
\noindent where the option {\tt "-fno-automatic"} indicates the static allocation 
of memory. The level-3 optimization ({\tt "-O3"}) should be switched on since it
improves the performance of the code considerably. For more details, and several
examples of compiling the program on different platforms, refer to Paper~III,
\S\,\refcompcomp.


\subsection{Input files}

The essential feature of the input data format is that there is only a very
\index{Input!general scheme}
short standard input file, which specifies: 
(i) the very basic parameters, such as $T_{\rm eff}$, $\log g$,
chemical abundances, etc.,
for which no reasonable default values can be specified;
(ii) the name of the file where the optional {\em keyword parameters} are set up;
and (iii) the names of files where the atomic data for the individual ions are
stored. Keyword parameters are defined as those for which the program assigns
default values, which are optimum for most applications, but could be changed
as required for a particular special application. They also enable one to choose among
several alternative numerical schemes, or to cope with convergence problems.
There are also several keyword parameters that determine the
overall computational strategy. The most important ones are described in 
Paper III, Chap.\,\refnonst, and the remaining ones  in Chaps.\,\refnsttwophys \  
and \refnsttwonum, and the full list of keyword parameters together with their default 
values is presented in Paper~III, Chap.\,\reflist.\\

Here is a list of basic input files.
\begin{itemize}
\item{\tt fort.1} --- The basic control file, containing just one single number,
specifying whether one calculates a stellar atmosphere or an accretion disk model.
If this number is 0, or if the file is  missing altogether, a
stellar atmosphere model is to be computed.
Otherwise, a disk vertical structure is computed.
\item{Standard input file (unit 5)} --- Main control data. It is a short file with only 
the most important parameters, and filenames of other files.
The structure of the file and the meaning of the individual input parameters
are explained in Paper III, Chap.\,\refnew.
\item{A file that specifies non-default values of the keyword parameters}.
Its name is specified in the standard input file.
\item{\tt fort.8} --- A starting model atmosphere in the case the calculation does not start
from the scratch (that is, with an LTE-gray model) - see Paper~III, Chap.\,\refmodin.
\end{itemize}
In the standard mode, where the opacities are computed on the fly, and
where one introduces the concept of explicit ions and levels (which is
mandatory for NLTE models), there is an important set of input files, namely
\begin{itemize}
\item{Files containing atomic data for the individual ions.}
These files are described in detail in Paper III, Chap.\,\refions. 
A collection of such files is available in the standard distribution of {\sc tlusty},
and in the {\sc tlusty} website.
\end{itemize}
There is another mode which uses pre-calculated opacity tables; for
more details and examples, refer to Paper~III, \S\,\refexampoptab\  and
\refnsttwooptab. In that case, the atomic data files are not needed.

There may be many more additional input files for specific applications, such
as white dwarfs, cools stars, etc. They are described in Paper III, Chap.\,\refnsttwophys.

All the input files are ASCII files to enable easy portability.
All the {\tt READ} statements use a free format. Moreover, Unit 5
may contain comment lines; {\sc tlusty} understands a line beginning
with {\tt *} or {\tt !} as a comment.


\subsection{Output files}

There are several output files. 
By default, all the output files are generated as ASCII files for portability.
{\sc tlusty} does not contain any explicit {\tt OPEN} statements for the
output files, so the files are generated with names {\tt fort.}{\it nn},
where {\it nn} is the corresponding unit number.

Here we only list the basic output files; there are more files used for
special purposes only. All the output files are described in detail in
Paper III, Chap.\,\refout.
\begin{itemize}
  \item Unit 6 -- Standard output
  \item {\tt fort.7} -- Condensed model atmosphere
  \item {\tt fort.9} -- Convergence log
  \item {\tt fort.10} -- Performance and error log
  \item {\tt fort.11} -- Mean opacities and other physical quantities for the resulting model
  \item {\tt fort.12} -- $b$-factors (NLTE departure coefficients) -- only for NLTE models             
  \item {\tt fort.13} -- Emergent flux in all frequency points (spectral energy distribution)
  \item {\tt fort.69} -- Timing log        
\end{itemize}


\subsection{Producing a model atmosphere with {\sc tlusty}}

Generically, the program can be run as follows:
\begin{verbatim}
        tlusty.exe < [std. input file] > [std. output file]
\end{verbatim}
A standard input file is mandatory. If the specification of the standard output
file is missing, the output is directed to the screen. It is however advisable
to save the standard output to a file because it contains a useful information
about the performance of the code, and will help in case of problems.

Although the file names can be arbitrary, it is advantageous to use the following convention: 
Any name of an input or output file is composed of a {\em core name} that may label the
basic physical parameters of a model, with an extension identifying the unit number.
For example, let us take a H-He model for $T_{\rm eff} = 35,000$ K and $\log g = 4$, in
LTE. Let the core name be {\tt hhe35lt}, so the standard input file is then {\tt hhe35lt.5}.
This file is a part of the standard distribution of {\sc tlusty}.

The code can then be run as follows
\begin{verbatim}
        tlusty.exe < hhe35lt.5  > hhe35lt.6
\end{verbatim}
where the standard output is redirected to the file {\tt hhe25lt.6} for
further inspection. However, in this case the executable file {\tt tlusty.exe}
has to be present in (or liked to) the current directory, and also the
necessary atomic data, in this case the files {\tt ./data/h1.dat}, {\tt ./data/he1.dat}, 
and {\tt ./data/he2.dat} have to be available. Therefore, one has to either create
a subdirectory {\tt ./data} in the current directory, or to set an appropriate
link to a universal directory that contains a collection of atomic data files.

An easier and safer way is to use a shell script {\tt RTlusty}, which is based
on a script {\tt XTlusty} kindly provided by Knox Long, and 
which is also a part of the standard distribution of {\sc tlusty}. It is called 
with one or two parameters,
\begin{verbatim}
RTlusty    model_core_name    [core_name_of_starting model]
\end{verbatim}
The second parameter does not have to be present if the model is calculated
from scratch. 
The executive part of the script looks like (schematically)
\begin{verbatim}
MOD=$1
if [ -e $2.7 ]  cp $2.7 fort.8
ln -s $TLUSTY/data data
$TLUSTY/tlusty/tlusty.exe  < $MOD.5 > $MOD.6
cp fort.7 $MOD.7
cp fort.9 $MOD.9
cp fort.69 $MOD.69
cp fort.13 $MOD.13
\end{verbatim}
The second command checks for the existence of the file that contains the
starting model, and if it does exist, copies it to {\tt fort.8}. The third command
links the general directory containing the necessary atomic data to a temporary
directory that is being referred to in the input files, and the last part runs the code
and stores the most important output files to corresponding files.\\

\noindent {\bf Important notes}\\
The script {\tt RTlusty} can run a model in any directory, provided that the 
two following requirements are fulfilled:
\begin{itemize}
\item
One has to set up an environment variable {\tt TLUSTY} that specifies the main 
{\sc tlusty} directory. For instance, using the tar files downloaded from the
Arizona site, and assuming that the tarball is extracted in the home directory,
the main {\sc tlusty} directory is generated as~ {\tt $\widetilde{}$/tlusty205}, one sets
\begin{verbatim}
setenv TLUSTY ~/tlusty205
\end{verbatim}
%

\item
The universal directory containing atomic data is a subdirectory
of the main {\sc tlusty} directory, that is, \${\tt TLUSTY/data}
\end{itemize}
Since the script is located  in the main {\sc tlusty} directory, it is advantageous
to set a {\tt path}  pointing to that directory. Otherwise, one would either need
to copy the script {\tt RTlusty} to the current directory, or to call it as
\${\tt TLUSTY/RTlusty}. In the following examples, we will call it simply as
{\tt RTlusty}.


\subsection{Test-case model atmosphere}
\label{testc}

The actual test case using the {\tt hhe35lt} is run as shown above, i.e.,
\begin{verbatim}
cd $TLUSTY/examples/hhe
RTlusty hhe35lt 
\end{verbatim}
which runs the code and stores not only the standard output (unit 6)
but also other important output files, all with the same core name,
and with the suffix that corresponds to the unit number.

The only input files
are the standard input file {\tt hhe35lt.5} and three atomic data files {\tt h1.dat},
{\tt he1.dat}, and {\tt he2.dat}, for H, He I, and He II, respectively. 
The structure of 
the standard input file is described in detail in Paper III, Chap.\,\refexamphhe,
and the structure of the atomic data files in Paper III, Chap.\,\refions.
The keyword parameter
file is missing, so that no keywords are required. The staring model file is missing too,
because the model is computed from scratch and no starting model is needed. 

A continuation of the run to produce a final NLTE model is described in detail in 
Paper~III, \S\,\refexamphhe. Briefly, the final NLTE model would be obtained by
\begin{verbatim}
RTlusty hhe35nc hhe35lt
RTlusty hhe35nl hhe35nc
\end{verbatim}
As explained in Paper~III, it is advantageous to proceed in three steps:
first to compute an LTE model, then a NLTE model with continua only (NLTC),
and finally a NLTE model with lines (NLTL).  This strategy follows a standard
practice of constructing NLTE models, first suggested by Auer \& Mihalas (1969).
The test case presented here produces a model that is similar, although somewhat
more involved, that the early NLTE model atmospheres produced by the pioneers
of this field (Auer \& Mihalas 1970, 1972; Mihalas \& Auer 1970; Mihalas et al. 1975). 

Paper~III, Clap. 6, presents a number of additional test cases; 
their input data may be used as templates for producing similar models.
All the test cases can be done by applying the script {\tt Runtest}, which 
calls elementary script {\tt RTlusty} in the individual directories after making 
necessary links to additional atomic data files.
The user is therefore encouraged to issue the following commands:
\begin{verbatim}
cd $TLUSTY/examples
Runtest 
\end{verbatim}
The script issues messages about a completion of the individual models that are 
being computed, and 
if everything goes well, it generates appropriate models in all the five subdirectories
of the directory {\tt examples}. The user may then compare the newly created files
with  the template output files that are provided and named with an additional label
{\tt .orig}, e.g., {\tt hhe35lt.7.orig}. A description of physics and 
numerics behind these models is explained in  detail in Paper~III, Chap. 6.


\subsection{How can I tell that the model is well converged?}
\label{howconv}

This is actually a very important question, which is however sometimes
overlooked. It is therefore important to look at this issue in more detail.
It should be fully realized that a model is constructed by applying an {\em iteration}
process. There are several possibilities of which actual scheme to apply, or
even which actual flavor is going to be used. They are explained from the
theoretical point of view in Paper~II, and their practical choice and setup is
described at length in Paper~III. In the present context, it suffice to say that
the program contains a number of default options; therefore without specifying
any additional parameters the code performs according to its default settings.
The basic iteration scheme is the hybrid CL/ALI (Complete Linearization/Accelerated
Lambda Iteration) method, described in Paper~II, \S\,3.2.

The first, and the most important criterion of convergence is the magnitude 
of the relative changes of the components of the state vector. 
The latter is defined as a set of all structural parameters (temperature, various particle
number densities, and the mean intensities of radiation in discretized frequency points)
in a given discretized depth point. The necessary condition for convergence
is that the maximum relative change of all components of the state vector in all
depths is ``small", say $10^{-3}$, as shown for instance on the example below. However, this
is generally {\em not} a sufficient condition. The reason is that in some cases
the maximum relative change can be ``small", but the absolute error of the solution
can still be large. This is known as the ``false convergence'', discussed in a similar 
context of solving the radiative transfer equation with a dominant scattering term in Hubeny  
\& Mihalas (2014, \S\,13.2). A usual symptom of such a behavior is that while the relative
changes are decreasing from iteration to iteration, the maximum relative change
decreases only very slowly, showing a decrease by an order of magnitude in 10 or more
iterations. 
                                                             
After a completed run, it is therefore mandatory to inspect the convergence log
to make sure that all the relative changes of the components of the state
vector are sufficiently small. The standard distribution of {\sc tlusty} also
contains an IDL program {\tt pconv.pro} which plots the contents of the
convergence log. The program requires as an input three output files from
{\sc tlusty}, namely units 7, 9, and 69,
and assumes that the file names are constructed using the above convention,
that is the core names are the same. The program is called as
\begin{verbatim}
pconv,'hhe35lt'
\end{verbatim}
It can be called also as
\begin{verbatim}
pconv
\end{verbatim}
in which case the files are supposed to have generic names {\tt fort.*}.
If IDL is not available, the user is strongly encouraged to write his/her
own graphical interface routine to display the contents of the file {\tt fort.9}.
Our xperience showed that by inspecting the file solely by the eye one can easily
miss significant spikes in relative changes.

A supplementary, and quite stringent criterion of the convergence of a 
model is based on examining
the conservation of the total flux. To this end, one needs to inspect the last 
table of the standard output, in the present
case {\tt hhe35lt.6}, that summarizes the computed model atmosphere.
The last four columns of the last table show the computed total flux (radiative plus, if applicable,
convective), and the ratios of the radiative, convective, and computed total
fluxes with respect to the theoretical total flux, $\sigma T_{\rm eff}^4$,
respectively. Therefore, the values of the last column should be very close to
unity (up to at most 1\% departures from it); otherwise the model cannot be 
viewed as sufficiently accurate, even if the convergence log may show that 
the model is formally converged. Again, this would be a consequence of the 
false convergence of the iteration scheme.

More information on convergence properties, and an outline of various strategies
to cope with convergence problems, is covered in Paper~III, Chaps. 10 and 15.

%
\begin{figure}[h]
\begin{center}
\label{fig1}
\includegraphics[width=4in]{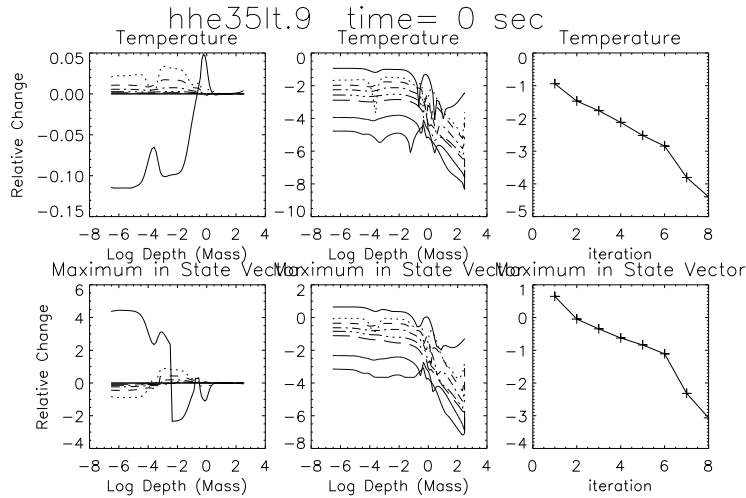}
\caption{Convergence log for a simple LTE H-He model.}
\end{center}
\vspace{-1em}
\end{figure}

Inspecting the results for the test model as suggested above shows that the
total flux is conserved within 0.1\% (the last table of {\tt hhe35lt.6}). Using the IDL
program {\tt pconv} to display the convergence log of {\tt hhe35lt.9}, see
Fig. 1, reveals an excellent and stable 
convergence behavior of the run (not surprisingly, because it is a very simple model).  
There are six panels on the figure, the upper three display the relative
changes of temperature (which is usually the most important component of
the state vector), while the lower three panels show the maximum relative change
of all components of the state vector. The leftmost panels show the relative change
as a function of column mass in the absolute scale, and the middle panels show the
same in the logarithmic scale. We display both, because the linear scale shows 
the sign of the relative changes but does not properly show those that are small.
These are in turn
clearly seen on the logarithmic scale. The rightmost panels show the maximum
relative change over all depth points. This is an indicator of the global convergence
of the model. Ideally, the relative changes should gradually decrease, which was
clearly the case here. A significant drop in the relative changes in the 7th
iteration originated due to the Ng acceleration (see Paper II, \S\,\refaccel). 
The figure also contains a header
that displays the core name and the total execution time. In the present case
it was less than 1\,s (actually 0.69\,s, as shown in {\tt hhe35lt.69}). This,
and all other calculations reported below, were
done on a MacBook Pro, with a 2.2.GHz Intel Core i7 CPU.


\section{Running {\sc synspec}}
\label{run_syn}
\subsection{Source code files and compilation}

Analogously to {\sc tlusty}, program {\sc synspec} is also distributed
as one big file together with four ``{\tt INCLUDE}'' files that declare
most of the arrays and specify their dimensions. The compilation
under {\tt gfortran} is done as
\begin{verbatim}
        gfortran -fno-automatic [-O3] -o synspec.exe synspec51.f
\end{verbatim}
%


\subsection{Input files}

The basic input files used for {\sc synspec} are exactly the same ones
as used for {\sc tlusty}, namely
\begin{itemize}
\index{Input!accretion disk switch}
\item{\tt fort.1} --- The basic control file, containing just one single number,
specifying whether one deals with a stellar atmosphere or an accretion disk model.
If this number is 0, or if the file is  missing altogether, a
stellar atmosphere model is considered,
otherwise, one ring of a disk model is considered.
\item{Standard input file (unit 5)} --- Main control data, exactly the
same as used for {\sc tlusty} to generate the model atmosphere for which
the synthetic spectrum is being calculated.
\item{File that specifies non-default values of the keyword parameters}.
Its name is specified in the standard input file. Again, this is the same file as that
used for running {\sc tlusty}. 
\item{\tt fort.8} --- Input model atmosphere  - typically given as previously computed
{\sc tlusty} model; more specifically, Unit 7 output from {\sc tlusty}, but it can also be a
Kurucz model.
\end{itemize}
Many of the input parameters do not have any meaning for running {\sc synspec},
but we kept the same input files for convenience. If one computes a synthetic
spectrum for a Kurucz model, one has to construct an artificial standard input file
for this run. It will be described briefly in \S\,\ref{run_kur}.

In addition to these files, there are several  input files specific to {\sc synspec}:
\begin{itemize}
\item {\tt fort.19} - atomic line list, based on Kurucz line lists. It will be described below
in \S\,\ref{linelist}.
\item {\tt fort.55} - additional input parameters for computing the synthetic spectrum.
It will  be described below in \S\,\ref{fort55}.
\item {\tt fort.56} - optional file that specifies a change of chemical abundances with
respect to the input model atmosphere.  It is not needed for LTE models or for NLTE 
models in which the chemical abundances considered for the {\sc synspec} run are
the same as those used in constructing the input model atmosphere. The file is
described below in \S\,\ref{fort56}.
\item {\tt fort.57} - energy bands for superlevels. This file is needed only if the input
model atmosphere is a NLTE metal-line blanketed model that treats the level structure
of the iron-peak elements in terms of superlevels and superlines. It is described in
Appendix B. We note that in the previous versions of {\sc synspec}, this file was called 
{\tt fort.13}. We changed the notation because using {\tt fort.13} as an input
file for {\sc synspec} could lead to a confusion
with another file {\tt fort.13} generated by {\sc tlusty} (a table of
emergent flux) which may accidentally be present in the working directory.
\end{itemize}
It should be stressed that while the atomic data used for {\sc tlusty} contain
sufficient information for evaluating the corresponding cross sections and opacities 
for bound-free transitions (continua), as well as bound-bound transition (lines),
{\sc synspec} only takes information for bound-free transitions from the
{\sc tlusty} input files, while the data for lines are taken from an
independent, {\sc synspec}-specific
line list. Typically, a line list used  by {\sc synspec} contains ``all'' lines, i.e. also data 
for lines of the species that were not considered by {\sc tlusty}.

Additionally, for some special purposes there are additional input files used only 
for some special purposes; these files are
also used by {\sc tlusty} for analogous purposes (white dwarfs; cool stars, etc.).
They must have specific names listed below. They are only used if a corresponding
parameter is set up -- see Appendix~A.
\begin{itemize}
\item  {\tt lemke.dat}, or {\tt tremblay.dat}, or {\tt hydprf.dat} - special hydrogen line 
broadening tables, see Appendix A, and, in the context
of {\sc tlusty}, Paper III, \S\,\refnsttwohyd 
\item {\tt laquasi.dat}, {\tt lbquasi.dat}, {\tt lgquasi.dat}, {\tt lhquasi.dat} - 
special hydrogen quasi-molecular data  - see Appendix A
\item{\tt tsuji.molec} -- a table with necessary parameters for the molecular state equation -- see Paper III, \S\,\refexampoptab, \S\,\refnonstglob\ and \S\,\refnstexaoptab
\item{\tt irwin.dat} -- a table of Irwin partition functions
-- see Paper III, \S\,\refnsttwoeos 
\end{itemize}
There are some other optional {\sc synspec}-specific input files, not being used by
{\sc tlusty}, 
\begin{itemize}
\item {\tt he1prf.dat}, {\tt he2prf.dat} -
special tables for He I and He II line broadening data - see Appendix A 
\end{itemize}
\noindent
Finally, if required, there are one or more molecular line lists
\begin{itemize}
\item {\tt fort.20} - basic molecular line list, containing data for most important
diatomic molecules: H${}_2$, C${}_2$, CH, CN, CO, 
 N${}_2$, NH, NO, O${}_2$, OH, MgH, MgC, MgN, MgO, SiH, SiC, SiN, SiO;
\item {\tt fort.21}, etc. - additional molecular line lists
\end{itemize}
This option is introduced in order to avoid working with excessively large files.
For instance, in addition to the line list for basic diatomic molecules, one may have
a special list for H$_2$O, for TiO and VO, and possibly more.

All the universal data files mentioned above are invoked by {\sc synspec} by appropriate
{\tt OPEN} statements, where they are assumed to reside in the directory {\tt ./data}.
Specifically, these are:
\begin{verbatim}
./data/lemke.dat, ./data/tremblay.dat, ./data/hydprf.dat, .
./data/he1prf.dat, ./data/he2prf.dat,
./data/laquasi.dat, ./data/lbquasi.dat, ./data/lgquasi.dat,
./data/lhquasi.dat,
./data/tsuji.molec, ./data/irwin.dat
\end{verbatim}
This means that by linking the general; ``{\tt data}" directory to {\tt ./data} as was
done before,
\begin{verbatim}
ln -s $TLUSTY/data data
\end{verbatim}
then all these data files are automatically available, in contrast to the previous
versions of {\sc synspec} where all the necessary links had to be set separately.
 

\subsection{Description of {\sc synspec}-specific input files}
\label{inpsyn}

\subsubsection{File  {\tt fort.55}}  
\label{fort55}
The general structure of the file, using the variable names exactly
as they are in the {\sc synspec} source code, is as follows:
\begin{verbatim}
  imode   idstd   iprin
  inmod   intrpl  ichang  ichemc
  iophli  nunalp  nunbet  nungam  nunbal
  ifreq   inlte   icontl  inlist  ifhe2
  ihydpr  ihe1pr  ihe2pr
  alam0   alast   cutof0  cutofs  relop  space
  nmlist, (iunitm(i),i=1,nmlist)
  vtb
  nmu0    ang0    iflux
\end{verbatim}
The parameters are read in a free format, and
the last three lines of the file do not have to be present if they are not needed. 
In that case, {\tt nmlist} and {\tt nmu0} are set to 0, and {\tt vtb} - the turbulent velocity -
is given by the {\sc tlusty} input.

Here is a brief explanation of the parameters:\\ [2pt]
$\bullet$ {\tt imode} -- sets the basic mode of operation:\\ 
= 0 - normal synthetic spectrum;\\
= 1 - spectrum for a few lines (obsolete);\\ 
= 2 - continuum (plus H and He II lines) only;\\ 
= 10 - spectrum with molecular lines.\\ 
= -1  an ``iron curtain option''. In this case one only computes
the opacity at the standard depth {\tt idstd} (see below), but does not solve the transfer equation. 
It is usually used with an artificial input model  atmosphere consisting of one single depth point.\\ [2pt]
$\bullet$ {\tt idstd}  - index of the ``standard depth'', which is defined as a depth where 
$T\approx (2/3) T_{\rm eff}$.
Approximately, ${\tt idstd} = (2/3) N\!D$, with $N\!D$ being the total number of depth points of the model. This parameter is not critical; it only influences the selection of lines 
to be considered, set through the rejection parameter {\tt relop} -- see Appendix C;\\ [2pt]
$\bullet$ {\tt iprin} -- when $>0$ increases the amount of printed information in the standard
output (essentially obsolete);\\ [2pt]
$\bullet$ {\tt inmod} -- an indicator of the input model:\\
$=0$ - the input model is a Kurucz model;\\ 
$=1$ - the input is a {\sc tlusty} model;\\ 
$=2$ - the input model is an accretion disk model \\ [2pt]
$\bullet$ {\tt intrpl, ichang} -- set the change of the input model (rarely used);\\ [2pt]
$\bullet$ {\tt ichemc} - if non-zero, indicates a change of abundances with respect 
to the input model. In this case, file {\tt fort.56} is required;\\ [2pt]
$\bullet$ {\tt iophli} - treatment of far L$\alpha$ wings (obsolete); should be set to 0;\\ [2pt]
$\bullet$ {\tt nunalp, nunbet, nungam, nunbal} -- if any of them set to a non-zero
value, the quasi-molecular
satellites of L$\alpha$, L$\beta$, L$\gamma$, and H$\alpha$, respectively, are
considered. In that case, additional input files containing the corresponding data
are needed. These have to have the following names: {\tt laquasi.dat}, {\tt lbquasi.dat}, 
{\tt lgquasi.dat}, and {\tt lhquasi.dat}, respectively. They are distributed along 
with {\sc synspec}. \\ [2pt]
$\bullet$ {\tt ifreq} -- the choice of  the radiative transfer equation solver; namely
if  {\tt ifreq} $\geq 10$ one uses the Feautrier scheme, otherwise the DFE scheme --
see Paper II. This option is rarely used; one usually uses the default value
{\tt ifreq} = 1.\\ [2pt]
$\bullet$ {\tt inlte} 
an auxiliary NLTE switch:\\
$= 0$ -  enforces LTE: lines are treated in LTE even if the input model is NLTE.\\
$> 0$ - its actual value (1 or 2) sets the mode of evaluation of the 
populations of levels close to the ionization limit - see \S\,\ref{what};\\ [2pt]
$\bullet$ {\tt icontl, inlist} -- obsolete parameters, typically set to 0; \\ [2pt]
$\bullet$ {\tt ifhe2} -- if set to a non-zero value, He II is treated as a hydrogenic
ion (see Appendix A);\\ [2pt]
$\bullet$ {\tt ihydpr, ihe1pr, ihe2pr} -- if set to a non-zero value, these parameters invoke a
special evaluation of line profiles of the respective atom/ion.  In this case, additional 
input files with corresponding data are needed -- see Appendix A;\\ [2pt]
$\bullet$ {\tt alam0} -- starting wavelength [\AA];\\ [2pt]
$\bullet$ {\tt alast} -- absolute value is ending wavelength [\AA],\\
if {\tt alast}$< 0$, then all the line wavelengths are in vacuum (notice that by default,
the wavelengths for $\lambda\leq 2000$ \AA\  are in vacuum, while for
$\lambda > 2000$ \AA\ they are in the air); \\ [2pt]
$\bullet$ {\tt cutof0} -- cutoff parameter: the opacity of any line except H and He II is cut at
{\tt cutof0} \AA\ from its center;\\ [2pt]
$\bullet$ {\tt cutofs} -- dummy variable; \\ [2pt]
$\bullet$ {\tt relop} -- line rejection parameter: a line is rejected
if the ratio of its line-center opacity to the continuum opacity at standard depth {\tt idstd}
is less than {\tt relop};\\ [2pt]
$\bullet$ {\tt space} -- spacing parameter,
which represents the maximum distance of two neighboring wavelength points
at the midpoint of the considered wavelength interval,
that is, $\Delta\lambda \leq {\tt space}$.
The actual maximum spacing is proportional to the wavelength,
$\Delta\lambda \leq {\tt space}\times ({\tt alam0} + {\tt alam1})/(2\lambda)$.
 For more details, see Appendix C.\\ [2pt]
$\bullet$ {\tt nmlist} -- number of additional molecular line lists, \\ [2pt]
$\bullet$ {\tt iunitm(i),i=1,nmlist} -- unit numbers of the additional molecular line lists
$\bullet$ {\tt vtb}  - if set, it specifies the turbulent velocity [km s${}^{-1}$]. This parameter
is only needed if the turbulent velocity for the current run of {\sc synspec} is  different 
from the value specified in the {\sc tlusty} input;\\ [2pt]
$\bullet$ {\tt nmu} -- if set to a non-zero value, {\sc synspec} produces not only a flux, but
also the specific intensities, $I_\nu(\mu)$ [erg cm${}^{-2}$s${}^{-1}$Hz${}^{-1}$ster${}^{-1}$];
$\mu$ is the cosine of the angle with respect to the normal to the surface. The value of {\tt nmu}
represents the number of the angle points, expressed as $\mu$, and set equidistantly
between {\tt ang0} and 1;\\ [2pt]
$\bullet$ {\tt ang0} -- minimum angle cosine;\\ [2pt]
$\bullet$ {\tt iflux} -- should be set to 1 to invoke an evaluation of specific intensities;\\ [2pt]

\subsubsection{File  {\tt fort.56}}  
\label{fort56}
File {\tt fort.56} has a simple structure:
\begin{verbatim}
      nchang
\end{verbatim} 
and then {\tt nchang} records, each with
\begin{verbatim}
      iatom  abn
\end{verbatim} 
where\\
$\bullet$ {\tt nchang} -- number of chemical elements for which  the chemical
abundances are changed with respect to the {\sc tlusty} input (specified in the
standard input file); \\ [2pt]
$\bullet$ {\tt iatom} -- atomic number;  \\ [2pt]
$\bullet$ {\tt abn} -- the modified abundance. The convention is as follows:

{\tt abn} $>0$ -- abundance by number, relative to hydrogen;

{\tt abn} $=0$ -- abundance is solar;

{\tt abn} $<0$ -- the absolute value is the ratio of the abundance to the solar one.

\medskip
For all so specified chemical elements, the values  of {\tt abn} represent 
the current chemical abundances for the run of {\sc synspec}.
For explicit, NLTE, species, they are also used
to multiply the NLTE level populations given by the model atmosphere input
by the ratio of the ``new" abundance of the corresponding chemical element
(specified by the values in the file {\tt fort.56}), and the ``old" abundance, referred
to in the standard input file, that were used previously by {\sc tlusty}. 
This procedure is not rigorous because the NLTE line formation is not a linear
process, and the modified abundances may yield a different global atmospheric 
structure. But this is acceptable if the abundance change concerns a minor
(not very abundant) chemical species. or for a major species if
the relative difference of the old and new
abundances is small, say, no more than 0.2 to 0.3 dex. For larger abundance
differences, it is more prudent first to recompute the model atmosphere and then
to produce a synthetic spectrum.

\subsubsection{Line list}
\label{linelist}

The line list contains one input record for each spectral line. In fact, it could
contain more records, but this option is cumbersome and is rarely used.
The input record contains the following variables:
\begin{verbatim}
alam anum gf excl ql excu qu agam gs gw inext
\end{verbatim}
and if the continuation indicator {\tt inext} is non-zero, then the next record is
\begin{verbatim}
wgr1 wgr2 wgr3 wgr4 ilwn iun iprf
\end{verbatim}
The meaning of these quantities is the following:\\ [2pt]
$\bullet$ {\tt alam} -- wavelength [nm]. One follows the convention that for
$\lambda < 200$~nm the wavelengths are in vacuum, while for $\lambda \geq 200$ nm the
wavelengths are in the air (they can however be modified to vacuum wavelengths by setting 
the parameter {\tt alast} in {\tt fort.55}---see above---to a negative value); \\ [1pt]
$\bullet$ {\tt anum} -- numerical code of the element and ion (using the Kurucz 
convention; e.g., 2.00 = He I, 6.03 = C IV, 26.01 Fe II, etc.)\\ [1pt]
$\bullet$ {\tt gf} -- $\log g\!f$ \\ [1pt]
$\bullet$ {\tt excl} -- excitation potential of the lower level [cm${}^{-1}$]\\ [1pt]
$\bullet$ {\tt ql} -- the $J$ quantum number of the lower level\\ [1pt]
$\bullet$ {\tt excu, qu} -- analogous quantities for the upper level\\ [1pt]
$\bullet$ {\tt agam} -- $\log\Gamma_{\rm rad}$ for radiation damping \\ [1pt]
$\bullet$ {\tt gs} -- $\log\Gamma_{\rm stark}$ for Stark broadening\\ [1pt]
$\bullet$ {\tt gw} -- $\log\Gamma_{\rm vdW}$ for Van der Waals broadening\\ [1pt]
$\bullet$ {\tt inext} -- if =1, next record is needed, where:\\ [4pt]
$\bullet$ {\tt wgr1 wgr2 wgr3 wgr4} -- Stark broadening values from Griem (1974)
tables, values for $T=5, 10, 20, 40 \times 10^3$ K, respectively\\ [1pt]
$\bullet$ {\tt ilwn} -- manual setting of the index of the lower level as 
used by {\sc tlusty}\\ [1pt]
$\bullet$ {\tt iun} -- manual setting of the index of the upper level as
used by {\sc tlusty}\\ [1pt]
$\bullet$ {\tt iprf} -- if non-zero, special procedure for evaluating the Stark broadening
(at present ony for He I -- see Appendix A).\\ 

In any of the parameters {\tt agam, gs, gw} is set to zero, one uses an approximate
formula for evaluating the corresponding broadening parameter. For details refer 
to Appendix A.


\subsection{Output files}
\label{syn_out}

\begin{itemize}
\item Unit 6 - standard output file. It contains a log of calculations, important
only in the case when something goes wrong.
  \item {\tt fort.7} -- synthetic spectrum. A simple table of wavelength [\AA] versus
  flux, expressed as the first moment of the specific intensity (Eddington flux),
  $H_\lambda$, in erg cm${}^{-2}$s${}^{-1}$\AA${}^{-1}$.
  \item {\tt fort.17} -- An analogous table of the theoretical continuum flux.
  \item {\tt fort.12} -- Identification table. It contains a list of all lines that were selected
  by {\sc synspec} to be taken into account, together with their basic parameters and
  an indicator of their approximate strength. This information can be used for an identification of the lines in the predicted spectrum. For more details, see Appendix D.
  \item {\tt fort.16} -- A list of partial equivalent widths of  {\sc synspec}-generated
  spectral regions -- see Appendix D.       
\end{itemize}


\subsection{Producing a synthetic spectrum for a Kurucz model}
\label{run_kur}

As pointed out above, {\sc synspec} accepts a Kurucz model structure
as an input model. One has to set the parameter {\tt inmod} in the file 
fort.55 (1st number on the second line) to 0 -- see \S\,\ref{fort55}.

The standard input to be used is essentially any input for {\sc tlusty}.
Recall that {\sc synspec} uses the standard input only to specify the
continuum opacity, that is the bound-free and free-free cross sections. Which
actual transitions are included is specified by the data. In hot stars where
the continuum opacity is dominated by H and He, it is actually sufficient
to use the sample file {\tt hhe35lt.5} as the standard input file for {\sc synspec}.
For cool stars, when the continuum opacity is dominated by H${}^-$ and
metal opacities, one has to construct the standard input accordingly.
The standard distribution of {\sc synspec} contains an example of such file, 
called {\tt cool.5}, which is safe to use
as the standard input to {\sc synspec} to produce synthetic spectra for
Kurucz models with $T_{\rm eff} \leq 10,000$ K.


\subsection{What chemical species and what lines are being considered?}
\label{what}

The choice of chemical species that are taken into account by {\sc synspec} is
controlled by the standard input data to {\sc tlusty}. As explained in more detail
in Paper III, \S\,\refnewat, the chemical elements can be treated either (i)
{\it explicitly} - treated in detail and in NLTE; (ii) {\it implicitly} - treated in LTE;
or (iii) rejected altogether. For each explicit element, a set of appropriate ionization
stages, called {\it explicit ions} is selected, and for each explicit ion a set of 
{\it explicit energy levels} is selected. This is specified in the atomic data file,
explained in detail in Paper III, Chap.\,\refions.

The continuum opacity sources are completely specified by the
standard input to {\sc tlusty}. As explained in detail in Paper III, \S\,\refnewion,
the continuum opacity is composed of all the bound-free transitions from explicit 
levels, and the free-free transitions of the explicit ions, plus possibly additional
opacity sources specified through the specific {\sc tlusty} input  parameters --
see Paper III, \S\,\refnsttwoopadd. The atomic level populations that enter
the evaluation of the corresponding opacities are taken directly from the 
{\sc tlusty} input model.

In contrast, lines are treated differently.
{\sc synspec} takes into account all lines of all
species that are not specifically rejected. For instance, in the test case shown in
\S\,\ref{syn_test}, H an He are treated explicitly; C, N, and O are treated 
implicitly, Li, Be, and B are rejected, and all the species with atomic number
larger than 8 are left unspecified. They are not taken into account by {\sc tlusty},
but since they are not explicitly rejected, all the lines of such species that appear
in the line list are being taken into account by {\sc synspec}. The rationale for
this approach is that when constructing a model atmosphere one takes into
account only the important species that may influence the global atmospheric
structure, while when constructing a synthetic spectrum using {\sc synspec} one is generally
interested in seeing all predicted lines. It should also be stressed that while the
atomic data files used by {\sc tlusty} contain data for selected lines, {\sc synspec}
takes a completely independent source of line data, namely the line list --
see \S\,\ref{linelist}.

\medskip

The atomic level populations that enter the evaluation of the corresponding 
opacity are determined in the following way:

(i) For non-explicit chemical elements, the level populations 
are evaluated in LTE. Obviously, if a model atmosphere is computed in LTE,
all level populations of non-rejected species (explicit and implicit) are evaluated
in LTE.

(ii) For explicit species in a NLTE model atmosphere, 
the level populations used for evaluating the bound-free
are free-free transitions are taken from the model atmosphere generated
by {\sc tlusty}. 
As specified above, the line list contains, for a given line, the $J$-values
and level energies for the lower and the upper level of the corresponding transition.
Based on this information, {\sc synspec} makes an association of the levels
being specified in the line list and the explicit levels that are specified through the
{\sc tlusty} input atomic data files.  This is described in detail in Appendix B.


\subsection{Test cases}
\label{syn_test}

Analogously to the script {\tt RTlusty}, we have constructed a similar script
{\tt RSynspec} that makes the necessary links, and run {\sc synspec}. 
It resides in the main {\sc tlusty} directory, \${\tt TLUSTY}.
The script is called with one or two or three parameters,
\begin{verbatim}
RSynspec core_name_of_input_model  file_fort.55  line_list
\end{verbatim}
The second parameter is the name of the file that is going to be linked to {\tt fort.55};
if it is missing, it is assumed that the file {\tt fort.55} exists in the current directory.
The third parameter is the filename of the line list; again, if it is missing, file
{\tt fort.19} already exists in the current directory.

The script stores the important output files as follows:
\begin{verbatim}
   cp fort.7 $MOD.spec
   cp fort.17 $MOD.cont
   cp fort.12 $MOD.iden
   cp fort.16 $MOD.eqws
\end{verbatim}
which represent the full spectrum continuum flux, identification table, and the
equivalent width table, respectively. 

\medskip

The first test case is computing a
synthetic spectrum for a short interval of wavelengths between 1400 and 1410~\AA,
for the model {\tt hhe35lt},  which is obtained as follows:
\begin{verbatim}
cd $TLUSTY/synspec/examples/hhe
RSynspec  hhe35lt  fort.55.lin   data/linelist.test
\end{verbatim}
where the input file {\tt fort.55.lin} has the following content
\begin{verbatim}
       0      50       0
       1       0       0       0
       0       0       0       0       0
       1       1       0       0       0
       0       0       0
    1400    1410      10       0  0.0001    0.01
       0       0
\end{verbatim}
The structure of the file is explained in \S\,\ref{fort55}. The 6th line
sets the parameters of the synthetic spectrum - first two numbers
of the limiting wavelengths of the interval to be covered, and the last
number the maximum spacing of the individual wavelength points,
all in \AA.

The standard distribution of {\sc synspec} also contains a sample of a line list 
used for tests, called {\tt linelist.test}, which has to be copied or linked to {\tt fort.19}.
The contents of {\tt fort.7}, the synthetic spectrum, and {\tt fort.17}, the theoretical
continuum, are displayed in Fig.2.
\begin{figure}[h]
\begin{center}
\label{fig2}
\includegraphics[width=4in]{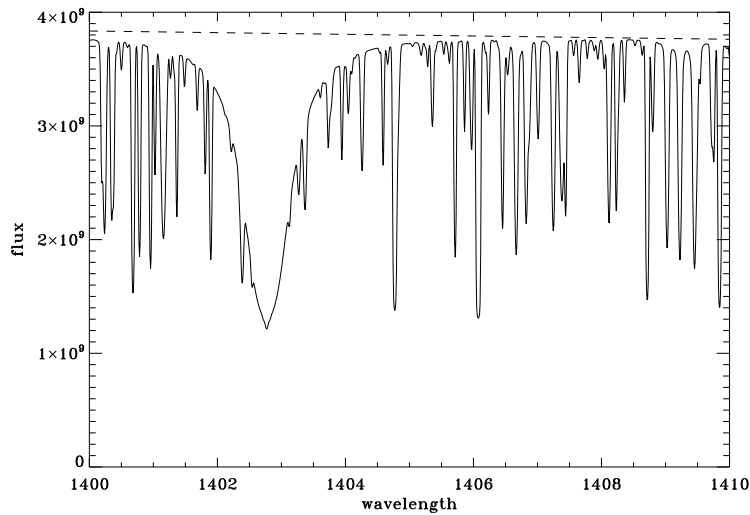}
\caption{Example of a synthetic spectrum for a simple LTE H-He model.
Dashed line portrays the theoretical continuum. The wavelength is in \AA,
and the flux is represented by the first moment of the specific intensity
$H_\lambda$ in erg~cm${}^{-2}$s${}^{-1}$\AA${}^{-1}$.}
\end{center}
\vspace{-1em}
\end{figure}

Another example is the predicted continuum spectrum in the optical region.
{\sc synspec} defines the continuum as the true continuum plus hydrogen and 
hydrogenic (in this case He II) lines. File {\tt fort.55}  (stored at {\tt fort.55.con}) is now:
\begin{verbatim}
       2      50       0
       1       0       0       0
       0       0       0       0       0
       1       1       0       0       0
       0       0       0
    3500    5000      10       0  0.0001    0.5
       0       0
\end{verbatim}
The continuum-only mode is set by the first number on the first line,  ${\tt imode}=2$.
In this case, no line list is needed.
The wavelength range is now set to 3500 to 5000 \AA, and the separation of the
wavelength points in the middle of the interval is 0.5 \AA; the wavelength spacing is 
proportional to the wavelength. 
The resulting synthetic spectrum is shown in Fig.3.
\begin{figure}[h]
\begin{center}
\label{fig2}
\includegraphics[width=4in]{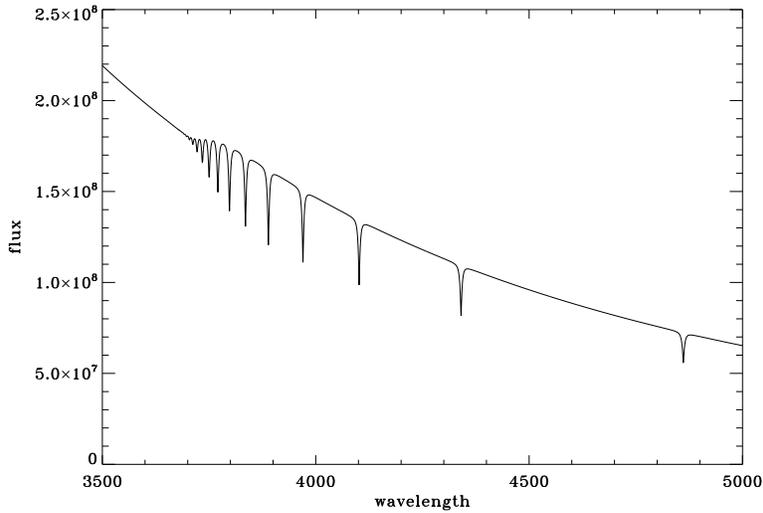}
\caption{Example of a synthetic spectrum for a simple LTE H-He model,
analogous to Fig. 2, but for the optical region and showing only continuum
plus H and He II lines. In this particular case, the H and He II lines have
a very similar strength.
The wavelength is in \AA,
and the flux is represented by the first moment of the specific intensity
$H_\lambda$ in erg~cm${}^{-2}$s${}^{-1}$\AA${}^{-1}$.}
\end{center}
\vspace{-1em}
\end{figure}

The next test case is a synthetic spectrum, in the same wavelength interval,
for a Kurucz model with $T_{\rm eff} = 9750$ K, $\log g =4$. The necessary input
files are located in directory \${\tt TLUSTY/synspec/examples/kurucz}. A file
with the model
atmosphere structure, analogous to {\sc tlusty} file {\tt fort.7} but in the specific Kurucz
format, is called {\tt ap00t9750g40k2.dat}. The standard input file, constructed
such as to provide appropriate continuum opacities as explained in \S\,\ref{run_kur},
is called {\tt cool.5}. Because the file names do not correspond to the standard
{\sc tlusty} naming convention, one has to modify the filenames in order to be able 
to use the script {\tt RSynspec}. This is provided by the script {\tt Runtest}, located in
this directory. Without using scripts {\tt Runtest} and {\tt RSynspec}, the synthetic
spectrum can be computed as follows:
\begin{verbatim}
ln -s -f $TLUSTY/data data
ln -s -f fort.55.lin fort.55
ln -s -f $TLUSTY/data/linelist.test fort.19
cp ap00t9750g40k2.daf fort.8
$TLUSTY/synspec/synspec.exe <cool.5 >tmp.log
\end{verbatim}
and possibly store resulting files {\tt fort.7}, etc., with a permanent name,
as the script {\tt Runtest} did, using a core name {\tt kurucz9750g40}.


\section{Working with {\sc ostar2002} and {\sc bstar2006} grids}

Computing modern and sophisticated NLTE metal line-blanketed model
atmospheres is time consuming; one model can easily take several hours to 
few tens of hours of computer time on a present mid-class workstation or
a Mac laptop. 

In many instances, it is actually not necessary to produce new models
from scratch. We have constructed two extensive grids of 
NLTE fully metal-line blanketed model atmospheres 
for O-stars (Lanz \& Hubeny 2003), called {\sc ostar2002}, and an analogous
grid {\sc bstar2006} (Lanz \& Hubeny 2007) for early B-stars.

The intent was to provide more or less definitive grids of models in the context 
of 1D plane--parallel geometry, with hydrostatic and radiative equilibrium, and 
without any unnecessary numerical approximations. The essential limitation of 
these models is the quality and availability of atomic data which, despite
recent efforts (such as the \textit{Opacity Project}, \textit{OPAL}, or \textit{IRON Project}) 
are still incomplete (for instance, the lack of available  collisional excitation 
cross--sections for dipole--forbidden transitions of the iron--peak elements).

The \textsc{ostar2002} grid contains 
model atmospheres for  $T_{\mathrm{eff}}$ between 27,500 and 55,000 K with the
step of 2,500 K, 
$\log g$ between 4.75 and a value that corresponds to an approximate location
of the Eddington limit, with the step 0.25, and for 10 metallicities: 2, 1, 1/2, 1/5, 1/10, 
1/30, 1/50, 1/100, 1/1000,  and 0 times the solar metal composition, so that the
grid is useful for studies of typical environments of massive stars: the Galactic center, 
the Magellanic Clouds, blue compact dwarf galaxies like I~Zw-18, and galaxies at high redshifts.
Departures from LTE are allowed for the following species: 
H, He, C, N, O, Ne, Si, P, S, Fe, Ni, in all important stages of ionization. 
There are altogether over 1,000 (super)levels treated in NLTE,  
about $10^7$ lines, and about  250,000 frequency points to describe the spectrum.

The \textsc{bstar2006} grid is similar. It contains models for 
$T_{\mathrm{eff}}$ between 15,000 and 30,000 K with the step of 1000 K,  
$\log g$ values are set similarly to the \textsc{ostar2002} grid,
and for 6 metallicities: 2, 1, 1/2, 1/5, 1/10, and 0 times solar. The species treated 
in NLTE are the same as in \textsc{ostar2002}, adding Mg and Al, but removing Ni, which 
is less important for B stars. There are altogether about 1,130
(super)levels treated in NLTE, about $10^7$ lines, and about 400,000 frequency points.
The models for both grids are available online\footnote{http://tlusty.oca.eu}.

There are two types of available files: (i) model structure -- essentially
three files; the standard input, {\tt *.5}; the keyword parameter file {\tt nst} referred to
by the the standard input (the same for all models), and the resulting condensed 
model atmosphere, {\tt *.7}.
(ii) Synthetic spectra, but with a relatively low resolution of about 1 \AA. To obtain
a higher resolution synthetic spectrum, one has to run {\sc synspec}, as shown
in \S\,\ref{grid_syn}.

\subsection{Do we need to compute a model atmosphere from scratch?}

For the range of effective temperature covered by the grids, it is advantageous
not to construct a model from scratch, but rather use an existing model from the
grid and possibly modify it to suit the user's purposes. In the following, we describe several
useful procedures to use existing models to produce detailed spectra for the basic
parameters ($T_{\rm eff}$, $\log g$, chemical abundances), either for those covered
exactly in the grids, or for different, but close, values of these parameters. We shall
also describe several procedures for upgrading the available models  
by updating model atoms and/or adding new explicit chemical species.

\subsection{Producing detailed spectra for grid models}
\label{grid_syn}

The websites of the model grids already contain a synthetic spectrum, but only
with a relatively low wavelength resolution of about 1 \AA. For many purposes,
one needs synthetic spectra of much higher resolution, or for wavelength regions
not covered by the distributed files, or perhaps with somewhat different chemical
abundances than those considered in the model grids. For all these purposes,
{\sc synspec} provides the necessary tool.

Running {\sc synspec}  is actually very simple. Taking an example of a sample B-star
model that is contained \${\tt TLUSTY/examples/bstar}, and also in the specific
directory \${\tt  TLUSTY/synspec/examples/bstar},
namely {\tt BGA20000g400v2} (for $T_{\rm eff}=20,000\,{\rm K}, \log g=4, v_{\rm tb}= 2$ km/s,
and solar composition), one produces a synthetic spectrum by using the script {\tt RSynspec}
\begin{verbatim}
cd $TLUSTY/synspec/examples/bstar
[cp superlevels fort.57]
RSynspec  BGA20000g400v2  fort.55.sol  data/linelist.test
\end{verbatim}
As mentioned before, this script accomplishes a number of tasks:\\ [2pt]
$\bullet$ links the general "data" directory \${\tt TLUSTY/.data} to {\tt ./data}, \\ 
$\bullet$ links the auxiliary input file {\tt fort.55.sol} to {\tt fort.55}; \\ 
$\bullet$ links the sample line list \${\tt TLUSTY/data/linelist.test} to {\tt fort.19} (notice
that since the script makes the link to the "data" directory first, one can refer to this
file as {\tt data/linelist.test}; \\ 
$\bullet$ copies the file {\tt BGA20000g400v2.7} (input  model atmosphere) to  {\tt fort.8}; \\ 
$\bullet$ runs {\sc synspec}, and \\ 
$\bullet$ stores important output files.
\medskip

If the file {\tt superlevels} is not copied to {\tt fort.57}, the association of levels referred to
in the line list and in the {\sc tlusty} input is done analogously as for light elements, 
without taking into account the parity of the levels. If the file {\tt fort.57} exists, an
information contained in it allows for a more consistent association of levels -- see Appendix B.
The latter is obviously a better option.

In this example, {\tt fort.55.sol} is analogous to that used in \S\,\ref{fort55}, namely
\begin{verbatim}
       0      34       0
       1       0       0       0
       0       0       0       0       0
       1       1       0       0       0
       0       0       0
    1400    1410      10       0  0.0001    0.01
       0       0
\end{verbatim}
which computes a short interval of spectrum between 1400 and 1410 \AA,
with the same abundances as those used in computing the model structure,
i.e., solar. No input file {\tt fort.56} is necessary. 

Another example is an analogous  spectrum, where we
change abundances of
the iron peak elements  (Cr to Ni) to 0.5 solar, and for silicon to twice solar, one 
makes a slight modification of the {\tt fort.55.sol}, which is called  {\tt fort.55.nonsol}
\begin{verbatim}
       0      34       0
       1       0       0       1
       0       0       0       0       0
       1       1       0       0       0
       0       0       0
    1400    1410      10       0  0.0001    0.01
       0       0
\end{verbatim}
In this case. one needs an additional file, {\tt fort.56}, stored as {\tt fort.56.nonsol},
which looks like this
\begin{verbatim}
      6
     14    -2.
     24    -0.5
     25    -0.5
     26    -0.5
     27    -0.5
     28    -0.5
\end{verbatim}
where again the negative values for abundances signify the abundances represented
as ratios with respect to the solar ones.  This file has to be  copied or linked to
{\tt fort.56} before running {\sc synspec} or invoking the script {\tt RSynspec}.
\medskip

Finally, to obtain a synthetic spectrum that is directly comparable to observations,
one has to convolve the net emergent flux with a rotational velocity of
the star, and possibly with an instrumental profile of the spectrograph that produced
an observed spectrum to be analyzed. To this end, one employs an additional
utility program called {\sc rotin}, which is described in more detail in Appendix E.

\subsection{Using {\sc synplot}}
\label{synplot}

For the users who have access to the IDL language, running {\sc synspec}
can be much more easily done through an IDL interface program called
{\sc synplot}\footnote{http://aegis.as.arizona.edu/$\,\,\,\widetilde{}\,\,$hubeny/pub/synplot2.1.tar.gz}
which sets up the input data, runs {\sc synspec} and {\sc rotin}, and produces
a plot of the resulting synthetic spectrum possibly also accompanied by  line
identification labels, among other functions that it can perform. All that is
accomplished by setting simple keyword parameters. A detailed user's
guide is a part of the main distribution file. We will just mention that the
above mentioned run of {\sc synspec}, moreover with a subsequent convolution with
rotational velocity $v \sin i = 20$ km/s, and the Gaussian instrumental profile
with FWHM=0.2 \AA, can by accomplished by the following
simple IDL statement
\begin{verbatim}
synplot,atmos='BGA20000g400v2',wstart=1400,wend=1410, $
abund=[14,14,-2.,24,28,-0.5],vrot=20,fwhm=0.2
\end{verbatim}

\subsection{Running a B-star model}
\label{grid_run}

As pointed out above, constructing a fully line-blanketed NLTE model from
the scratch is not trivial, for one often needs to proceed in several judiciously
chosen intermediate steps -- for more on this topic, refer to Paper III, 
Chaps.\,\refstrateg\ and \reftrouble. Therefore, when one needs to construct 
a model within the range
of the {\sc bstar2006} or {\sc ostar2002} model grids, it is easier to start with the
closest existing model and use it as the starting model for the run. 

In this section, we describe this approach in the case where the choice of explicit
atoms, ions, and levels remains the same as in the original model. The case
when one needs to change the structure of explicit atoms, ions, or levels, is
described below in \S\,\ref{grid_addspe} and \S\,\ref{grid_addlev}.

As an example, let us assume that one intends to construct a model with
$T_{\rm eff}=20,500$ K, $\log g=4.1$, and with the iron abundance being 1.5
times the solar abundance, and the Mg and Si abundance 0.6 times solar. 
We set up a new standard input file and call it,
in analogy with the {\sc bstar2006} grid convention, as {\tt BGA21000g410sp.5}.
It is very similar to the original {\tt BGA20000g400v2.5} file. 
For convenience, we will show it below: \\ [2pt]

\hrule
\hrule
\begin{verbatim}
20500. 4.1         ! TEFF, GRAV 
 F  F              ! LTE,  LTGRAY
 'nst'             ! keyword parameters filename
*
* frequencies
*
 2000
*
* data for atoms   
*
 30                 ! NATOMS
* mode abn modpf
    2   0.      0  ! H
    2   0.      0  ! He
    0   0.      0
    0   0.      0
    0   0.      0
    2   0.      0  ! C
    2   0.      0  ! N
    2   0.      0  ! O
    1   0       0
    2   0.      0  ! Ne
    1   0.      0
    2  -0.6     0  ! Mg
    2   0.      0  ! Al
    2  -0.6     0  ! Si
    1   0.      0  ! P
    2   0.      0  ! S
    1   0.      0
    1   0.      0
    1   0.      0
    1   0.      0
    1   0.      0
    1   0.      0
    1   0.      0
    1   0.      0
    1   0.      0
    2  -1.5     0  ! Fe
    1   0.      0
    0   0.      0  
    1   0       0
    1   0       0
*
* data for ions
*
*iat   iz   nlevs  ilast ilvlin  nonstd typion  filei
*
   1    0     9    0    0    0   ' H 1' 'data/h1.dat'
   1    1     1    1    0    0   ' H 2' ' '
   2    0    24    0    0    0   'He 1' 'data/he1.dat'
   2    1    20    0    0    0   'He 2' 'data/he2.dat'
   2    2    1     1    0    0   'He 3' ' '
   6    0    40    0    0    0   ' C 1' 'data/c1.dat'
   6    1    22    0    0    0   ' C 2' 'data/c2.dat'
   6    2    46    0    0    0   ' C 3' 'data/c3_34+12lev.dat'
   6    3    25    0    0    0   ' C 4' 'data/c4.dat'
   6    4     1    1    0    0   ' C 5' ' '
   7    0    34    0    0    0   ' N 1' 'data/n1.dat'
   7    1    42    0    0    0   ' N 2' 'data/n2_32+10lev.dat'
   7    2    32    0    0    0   ' N 3' 'data/n3.dat'
   7    3    48    0    0    0   ' N 4' 'data/n4_34+14lev.dat'
   7    4    16    0    0    0   ' N 5' 'data/n5.dat'
   7    5     1    1    0    0   ' N 6' ' '
   8    0    33    0    0    0   ' O 1' 'data/o1_23+10lev.dat'
   8    1    48    0    0    0   ' O 2' 'data/o2_36+12lev.dat'
   8    2    41    0    0    0   ' O 3' 'data/o3_28+13lev.dat'
   8    3    39    0    0    0   ' O 4' 'data/o4.dat'
   8    4     6    0    0    0   ' O 5' 'data/o5.dat'
   8    5     1    1    0    0   ' O 6' ' '
  10    0    35    0    0    0   'Ne 1' 'data/ne1_23+12lev.dat'
  10    1    32    0    0    0   'Ne 2' 'data/ne2_23+9lev.dat'
  10    2    34    0    0    0   'Ne 3' 'data/ne3_22+12lev.dat'
  10    3    12    0    0    0   'Ne 4' 'data/ne4.dat'
  10    4     1    1    0    0   'Ne 5' ' '
  12    1    25    0    0    0   'Mg 2' 'data/mg2.dat'
  12    2    1     1    0    0   'Mg 3' ' '
  13    1    29    0    0    0   'Al 2' 'data/al2_20+9lev.dat'
  13    2    23    0    0    0   'Al 3' 'data/al3_19+4lev.dat'
  13    3    1     1    0    0   'Al 4' ' '
  14    1    40    0    0    0   'Si 2' 'data/si2_36+4lev.dat'
  14    2    30    0    0    0   'Si 3' 'data/si3.dat'
  14    3    23    0    0    0   'Si 4' 'data/si4.dat'
  14    4    1     1    0    0   'Si 5' ' '
  16    1    33    0    0    0   ' S 2' 'data/s2_23+10lev.dat'
  16    2    41    0    0    0   ' S 3' 'data/s3_29+12lev.dat'
  16    3    38    0    0    0   ' S 4' 'data/s4_33+5lev.dat'
  16    4    25    0    0    0   ' S 5' 'data/s5_20+5lev.dat'
  16    5    1     1    0    0   ' S 6' ' '
  26    1    36    0    0   -1   'Fe 2' 'data/fe2v.dat'
   0    0                               'data/gf2601.gam'
                                        'data/gf2601.lin'
                                        'data/fe2p_14+11lev.rap'
  26    2    50    0    0   -1   'Fe 3' 'data/fe3v.dat'
   0    0                               'data/gf2602.gam'
                                        'data/gf2602.lin'
                                        'data/fe3p_22+7lev.rap'
  26    3    43    0    0   -1   'Fe 4' 'data/fe4v.dat'
   0    0                               'data/gf2603.gam'
                                        'data/gf2603.lin'
                                        'data/fe4p_21+11lev.rap'
  26    4    42    0    0   -1   'Fe 5' 'data/fe5v.dat'
   0    0                               'data/gf2604.gam'
                                        'data/gf2604.lin'
                                        'data/fe5p_19+11lev.rap'
  26    5     1     1    0    0   'Fe 6' ' '
   0    0     0    -1    0    0  '    ' ' '
*
* end
\end{verbatim}
\hrule
\hrule
\bigskip

Notice that the only change with respect to {\tt BGA20000G400v2.5} is in
the first line that specifies the basic model parameters, and in the appropriate entries
for Mg, Si, and Fe, when the chosen non-solar abundances are specified.
The same convention for coding the chemical abundances as that shown in
\S\,\ref{fort55} is used here as well; that is, if a number is negative, its absolute value
expresses the ratio of the current abundance to the solar one.
Since the choice of explicit ions and levels remains the same as in the {\sc bstar2006}
grid models, the rest of the input file is unchanged. 

The structure of the input files, and the meaning of the individual parameters,
is explained in detail in the appropriate chapters of Paper~III. For the present purposes,
we mention that the first three entries of the input data for each ion represent the
atomic number of the parent species, charge, and the number of explicit levels,
and the last two are the label of the ion, and the filename of the atomic data file.
The highest ionization stage of each chemical element must be represented by a
one-level ion.

Before running {\sc tlusty}, one has to check that the keyword parameter file,
in this case called {\tt nst}, is present in the current directory, or is being properly
linked to the corresponding file located elsewhere. One can also make a link to
the atomic data files, or simply use the script {\tt RTlusty}, described in  \S\,\ref{testc}.
In this case, the model is run as
\begin{verbatim}
cd $TLUSTY/examples/bstar
RTlusty  BGA20500g410sp  BGA20000g400v2
\end{verbatim}
This run is not considered as a test case because to fully converge the model may take
several hours. This case is presented to provide a guidance to the user about
how to construct new models from the models available in the {\sc bstar2006}
grid.

\subsection{Interpolating model structures or predicted spectra?}
\label{grid_inter}

Since the model atmosphere grids have rather densely spaced values of the
basic parameters ($T_{\rm eff}$, $\log g$, and chemical abundances), it is
actually possible to avoid a rather time-consuming evaluation of a model
with different basic parameters, and instead to interpolate between models.
Taking the above example, the existing models that bracket the desired values
are {\tt BGA20000g400v2, BGA21000g400v2, BGA20000g425v2}, and 
{\tt BGA21000g425v2}. A slight complication arises due to requiring non-solar chemical
abundances of Mg, Si, and Fe. However, since they do not differ significantly
from the solar abundances, they can be neglected for constructing the model
structure, and be taken into account only in {\sc synspec}.

The question is now which of the two following procedures is more accurate:\\ [2pt]
(i) one can construct synthetic spectra for the four bracketing models specified
above, with appropriate Mg, Si, and Fe abundances, and then interpolate the resulting spectra to the desired values of $T_{\rm eff}$ and $\log g$, or \\ [2pt]
(ii) one interpolates the corresponding structural parameters (temperature, electron
density, all atomic level populations) from the bracketing models to the desired
values of $T_{\rm eff}$ and $\log g$, using a log-log interpolation, and then
compute a single synthetic spectrum for this new interpolated model.

This question was studied in Lanz \& Hubeny (2003), who have demonstrated that it is 
in most cases preferable to use the second option, i.e., interpolate the models 
and then to compute a synthetic spectrum. It is also faster, because one computes
a synthetic spectrum only once.

\subsection{Adding new NLTE species to existing models}
\label{grid_addspe}

For some applications, it might be desirable to add a completely new 
explicit (i.e., NLTE) chemical element to the set of the original explicit species.
We stress that this is an issue for NLTE models only, because for LTE
models the only independent structural parameters are the temperature,
the total particle number density, and the electron density, while the 
individual level populations are given as functions of these parameters.
For NLTE models, the individual level populations have to be stored
as independent state parameters, and therefore when adding new levels
or new species one has to make sure that the level populations taken from
the input model are properly assigned to a new system of explicit levels.

A relatively simple task is adding new NLTE species to existing models.
If the atomic level structure remains otherwise unchanged, this can be
accomplished easily. 
For example, let us assume that one intends to add
argon to the set of explicit species. One has of course first construct the
appropriate atomic data files. This was actually done (Lanz et al. 2008);
the corresponding files are also present in the {\sc tlusty} distribution.
The standard input for calculating such model, say for 
$T_{\rm eff}=20,000$ K and  $\log g=4.0$, and with solar abundances, 
is very similar to {\tt BGA20000g400v2.5}. 

We will call the new file
{\tt BGA20000t400ar.5}. Here is the structure of the first block of the file: \\ [2pt]

\hrule
\hrule
\begin{verbatim}
20000. 4.0         ! TEFF, GRAV 
 F  F              ! LTE,  LTGRAY
 'nst-ar'          ! keyword parameters filename
*
* frequencies
*
 2000
*
* data for atoms   
*
 30                 ! NATOMS
* mode abn modpf
    2   0.      0  ! H
    2   0.      0  ! He
    0   0.      0
    0   0.      0
    0   0.      0
    2   0.      0  ! C
    2   0.      0  ! N
    2   0.      0  ! O
    1   0       0
    2   0.      0  ! Ne
    1   0.      0
    1   0.      0  ! Mg
    1   0.      0  ! Al
    2   0.      0  ! Si
    2   0.      0  ! P
    2   0.      0  ! S
    1   0.      0
    2   0.      0  ! Ar
    1   0.      0
    1   0.      0
    1   0.      0
    1   0.      0
    1   0.      0
    1   0.      0
    1   0.      0
    2   0.      0  ! Fe
    1   0.      0
    0   0.      0  
    1   0       0
    1   0       0
*
\end{verbatim}
\hrule
\hrule
\bigskip
The only two changes with respect to {\tt BGA20000g400v2.5} is 
different keyword parameter file, {\tt nst-ar}, and a specification
of the argon being now considered as an explicit species.

Since the order of explicit ions is arbitrary, one can place the ions of argon
at the end of the file, just after the record for Fe 6, \\ [2pt]
\hrule
\hrule
\begin{verbatim}
  26    5     1    1    0    0   'Fe 6' ' '
  18    0    71    0    0    0   'Ar 1' 'data/ar1_71lev-jK.dat'
  18    1    54    0    0    0   'Ar 2' 'data/ar2_42+12lev.dat'
  18    2    44    0    0    0   'Ar 3' 'data/ar3_27+17lev.dat'
  18    3    26    0    0    0   'Ar 4' 'data/ar4_25+11lev.dat' 
  18    4     1    1    0    0   'Ar 5' ' '  
   0    0     0   -1    0    0   '    ' ' '
*
* end
\end{verbatim}
\hrule
\hrule
\bigskip

The keyword parameter file {\tt nst-ar} is the analogous to {\tt nst},
one only  adds a new keyword parameter, namely 
\begin{verbatim}
ICHANG=1
\end{verbatim}
which signals that the global atomic level structure was changed (energy
levels of Ar I to Ar IV were added), but that one deals with a simple change,
namely adding energy levels of the new species at the end. 
The populations of the explicit (NLTE) levels
except those of Ar are taken from the input model atmosphere, while the populations
of all the argon levels are initialized to their LTE values.
Again, this model is not considered as a test case because to fully converge the
tmodel may take several hours, but is presented to provide a guidance to the user about
how to proceed when adding new chemical elements to exisiting models.

\subsection{Adding atomic energy levels to existing models}
\label{grid_addlev}

In some cases, one may also want to update the existing atomic data file
by adding new explicit levels to an already selected explicit ion. For example,
one intends to perform a detailed study of neon line formation. The standard
data for Ne I and Ne II consider 35 and 32 explicit levels, respectively. One may
wish to use more sophisticated model atoms, as was done by Cunha et al (2006).
This study considered 79 explicit levels for Ne I, and 138 levels for Ne II. The
corresponding data files are also present in the {\sc tlusty} distribution. To run such
a model, one has first to modify the input records for Ne I and Ne II to
\begin{verbatim}
  10    0    79    0    0    0   'Ne 1' 'data/ne1_79lev-jK_c1.dat'
  10    1   138    0    0    0   'Ne 2' 'data/ne2_138lev.dat'
\end{verbatim}
One cannot place the new levels at the end of the file, so one has to communicate
to the code that the level structure is being changed in a complex way, and to
specify how to initialize the populations of the newly introduced explicit levels.
To this end, one has to set the keyword parameter in the {\tt nst} file as
\begin{verbatim}
ICHANG=-1
\end{verbatim}
Next, one has to specify an initialization of the level populations. To this end, one 
has to add NLEVEL records (NLEVEL being the total number of the explicit levels 
in the new model) at the end of the standard input file, each representing a mode
of evaluation of the initial level population.
It is done as follows (a more detailed description is presented in Paper III,
\S\,\refinpchan): \\ [2pt]
(i) all levels of hydrogen through  oxygen are unchanged, and their population is taken
from the input model atmosphere. Counting these levels, one finds that there are
$9+1+24+20+1+40+22+46+25+1+34+42+32+48+16+1+33+48+41+39+6+1=530$
such levels (see the listing of the file presented in \S\,\ref{grid_run}). Thus,
the first 530 records that are being placed at the end of the standard input file are:
\begin{verbatim}
    1    0    0    0    0    0    1.
    2    0    0    0    0    0    1.
    .....
  530    0    0    0    0    0    1.
\end{verbatim}
which specify that the level populations of the first 530 levels are initialized
as the first 530 level populations of the input model. \\ [2pt]
(ii) Next, one has to specify the initialization of the 79 levels of Ne I, so one has to
set 79 records for each such level. The initialization can be done in several 
different ways; for simplicity we will
assume here that all the populations of the Ne I levels except of the ground state
are initialized to LTE populations. In this case the input records are:
\begin{verbatim}
  531    0    0    0    0    0    1.
    0    1  566    0    0    0    1.
 .. and the same record 77 more times
\end{verbatim}
Here, the first record specifies that the ground state of Ne I, with the
global level index 531, is again taken from the input model, while the
next 78 levels (all excited states of Ne I) are initialized to LTE. Specifically,
the first number, =0, indicates that there is no equivalent level in the
original model\footnote{In fact, some newly specified levels can have
equivalents in the old system of levels, either having an identical counterpart,
or originating by a splitting of an original level composed of several
actual levels, etc. {\sc tlusty} can deal with such a situation, see Paper~III, \S\,\refinpchan,
but since this procedure influences only the initial level populations which are going to
be immediately recomputed by the code anyway, a simple procedure outlined here
is satisfactory.};
the second number, = 1, indicates LTE, an the third
number specifies that the LTE is taken with respect to the ground state
of the next ion (Ne II), which has a global index in the old model equal to 566
(i.e. 530 previous levels for H -- O, plus 35 levels in the old model atom for Ne I,
plus 1).

Analogously, for Ne II, one has altogether 138 records, again one for the ground state
and 137 identical ones for the excited states, 
\begin{verbatim}
  566    0    0    0    0    0    1.
    0    1  598    0    0    0    1.
 .. and the same record 136 more times
\end{verbatim}
because the ground state of Ne III has global index 598 in the input model.\\ [2pt]
(iii) The levels of the subsequent ions of Ne III - Ne V, and all ions of Mg, Al, Si, S, 
and Fe are unchanged.
There are altogether 
$34+12+1+25+1+29+23+1+40+30+23+1+33+41+38+25+1+36+50+43+42+1=
530$ such levels (which is just a coincidence
with the previous unmodified  530 levels). One will thus have 530 additional
records analogous to the first 530 records, namely
\begin{verbatim}
  598    0    0    0    0    0    1.
  599    0    0    0    0    0    1.
    .....
 1127    0    0    0    0    0    1.
\end{verbatim}
because the total number of all explicit levels in the input model is
$530+35+32+530=1127$.


\section{What is next?}

The present paper serves as a truly brief introduction to working with
{\sc tlusty},  
and provides only a rudimentary information
about the code. To gain a more thorough understanding of
the physics and numerics involved, the user is encouraged to study
Paper II. For understanding of the input parameters and many options
offered by {\sc tlusty}, Paper III provides the necessary information and guidance.

For {\sc synspec}, which is conceptually much simpler, this document provides
a complete guide for its operation. The only additional information relevant to
{\sc synspec} covered in Paper~II is a description of the available radiative transfer
equation solvers, and a description of the standard input file used by both {\sc tlusty}
and {\sc synspec}, given in Paper~III. 

\section* {Acknowledgements}
A more complete list of a large number of colleagues that contributed to the
development of {\sc tlusty} and {\sc synspec}, or who encouraged us to implement
various upgrades, is presented in Paper~III. Here we would like to gratefully
acknowledge a useful feedback on this paper provided by Yeisson Ossorio, 
Klaus Werner, Marcos Diaz, and, in particular, Peter Nemeth and Knox Long.
I.H. gratefully acknowledges the support from the Alexander von Humboldt Foundation,
and wishes to thank especially to Klaus Werner for his hospitality at the Institute of Astronomy 
and Astrophysics of the University of T\"ubingen, where a part of the work on this paper was done. 

\bigskip

\section*{Appendix A: Treatment of line profiles in {\sc synspec}}
\addcontentsline{toc}{section}{Appendix A: Treatment of line profiles in {\sc synspec}}

The treatment of line profiles in {\sc synspec} is independent of that used in 
{\sc tlusty}. The rationale for this approach is that in {\sc tlusty} the main emphasis
is on the global solution of the all structural equations, where often, in particular
for weak lines, details of the treatment of line profiles are inconsequential. 
Moreover, the computer time needed to compute a model may already be quite
large, so one can afford, or even should use, approximations  of the line profiles of 
weaker lines. In contrast, {\sc synspec} only computes an emergent radiation,
so this should be done as accurately as possible, and consequently an evaluation 
of intrinsic line profiles is more involved.
There is a different treatment of lines of H, He~I, He~II, and all the
other species.
\begin{itemize}
\item Hydrogen:
The lines of hydrogen are not included in the line list; in fact, the hydrogen lines are
treated as a part of the continuum. There are several options, controlled by the 
parameter {\tt ihydpr} from the input file {\tt fort.55}: 
\begin{itemize}
\item if ${\tt ihydpr}=0$, then an approximate evaluation of Doppler + Stark
broadening of hydrogen lines after Hubeny et al. (1994) is used.
\item  if ${\tt ihydpr}=1$, the profiles are computed using Lemke's (1997)
Stark broadening tables, contained in the file {\tt ./data/lemke.dat}. 
\item  if ${\tt ihydpr}=2$, the profiles are computed using Tremblay \& Bergeron
(2009) tables, contained in the file {\tt ./data/tremblay.dat}.
\item if ${\tt ihydpr} < 0$, the line profiles are computed using Schoening \& Butler (1989)
tables, contained in the file {\tt ./data/hydprf.dat}.
\end{itemize}
There is another possible modification of the hydrogen line profiles which can be
used together with any of the above options, namely including the
{\em quasi-molecular satellites} of the hydrogen Lyman lines and the
H$\alpha$ line, after Allard \& Koester (1992). Their inclusion
is controlled by the {\tt fort.55} parameters {\tt  nunalp, nunbet, nungam}, and {\tt nunbal}
for Lyman~$\alpha$, $\beta$, $\gamma$, and H$\alpha$ satellites, respectively.
If any of these are set to a non-zero value\footnote{Unlike previous versions of {\sc synspec}.
the actual values of these parameters do not have any specific meaning; the code now
uses explicit {\tt OPEN} statement(s) that specifies the filename(s) of the
file(s) that contain the corresponding table.}, an evaluation of the corresponding satellite
opacity is switched on. 
The data are contained in the files
{\tt ./data/laquasi.dat, ./data/lbquasi.dat, ./data/lgquasi.dat}, and {\tt ./data/lhquasi.dat}. 

\item He I: There are three options of treating these lines:
\begin{itemize}
\item If nothing else is specified, the lines are treated as any other metal lines, see
below.
\item If one sets the input parameter {\tt ihe1pr} in {\tt fort.55} (see \S\,\ref{fort55})
to a non-zero value, the four triplet lines, $\lambda\lambda$ 4026, 4387, 4471, and 4921~\AA,
are treated using special line broadening tables, after Barnard et al. (1974) and
Shamey (1969). The tables are contained in the file {\tt./data/he1prf.dat}.
\item A more cumbersome option is to change by hand the parameter {\tt iprf}
in the line list from the default value of 0 to a specific non-zero value. In this
case one will use the treatment of Dimitrijevic \& Sahal-Br\'echot (1984), with
the appropriate data already hardwired in the code. The association of the 
parameter {\tt iprf} and the line is the following:
\begin{center}
\begin{tabbing}
$\lambda$ (\AA) \ \ \ \ \ \  \=   {\tt iprf} \ \ \ \ \ \  \=  $\lambda$ (\AA) \ \ \ \ \ \  \=   {\tt iprf} \\ [2pt]
3819.60 \>  \ \ 1 \>  4437.55 \>  \ \ 11 \\
3867.50 \>  \ \ 2 \>  4471.50 \>  \ \ 12 \\
3871.79 \>  \ \ 3 \>  4713.20 \>  \ \ 13 \\
3888.65 \>  \ \ 4 \>  4921.93 \>  \ \ 14 \\
3926.53 \>  \ \ 5 \>  5015.68 \>  \ \ 15 \\
3964.73 \>  \ \ 6 \>  5047.74 \>  \ \ 16 \\
4009.27 \>  \ \ 7 \>  5875.70 \>  \ \ 17 \\
4120.80 \>  \ \ 8 \>  6678.15 \>  \ \ 18 \\
4120.80 \>  \ \ 9 \>  4026.20 \>  \ \ 19 \\
4168.97 \>  \ 10\>  4387.93 \>  \ \ 20 \\
\end{tabbing}
\end{center}

\end{itemize}

\item He II:
There are two switches that influence  the He II lines, both input through
the file {\tt fort.55}, {\tt ifhe2} and {\tt ihe2pr}. The parameter {\tt ifhe2} sets
a general treatment of the He II opacity:
\begin{itemize}
\item if ${\tt ifhe2} > 0$, He~II is treated as a hydrogenic ion, that is, its
lines are taken into account even if they do not appear in the line list. If
He~II data appear in the line list, they are disregarded and replaced by analytical
values for hydrogenic ions. This option is more exact, but is not adopted generally
because for cool temperatures, $T < 10^4$~K, the He~II ion has a very low
population which can sometimes lead to numerical problems.
\item  if ${\tt ifhe2} = 0$, the hydrogenic property of He~II is not taken into
account; the He II lines are included only if they appear in the line list, and
with parameters given by the line list.
\end{itemize}
Similarly to He I, there are several options to evaluate the line profiles, 
controlled by the parameter {\tt ihe2pr}:
\begin{itemize}
\item if ${\tt ihe2pr} = 0$, the He II lines are teated by the default way, i.e.\\
-- if {\tt ifhe2} $> 0$, using the approximate hydrogenic Stark + Doppler profile 
after Hubeny et al. (1994);\\
-- if {\tt ifhe2} $= 0$, they are treated as any other metal lines with the broadening 
parameters  specified in the line list.
\item if ${\tt ihe2pr} > 0$, the He II line profiles are given through the Stark
broadening tables of Schoening \& Butler (1989), contained in the file 
{\tt ./data/he2prf.dat}. 
\end{itemize} 

\item All other species:
Line profiles are given by the Voigt profile, $H(a,x)$, where 
$x= (\nu-\nu_0)/\Delta\nu_D$, and $a=\Gamma/(4\pi\Delta\nu_D)$.
Here, $\Delta\nu_D$ is the Doppler width, $\nu_0$ the line center frequency,
$a$ is the Voigt damping parameter, and $\Gamma$ is the physical damping 
parameter. The latter is given by
\begin{equation}
\Gamma = \Gamma_{\rm rad} + \Gamma_{\rm Stark} + \Gamma_{\rm vdW},
\end{equation}
where $\Gamma_{\rm rad}, \Gamma_{Stark}$ and $\Gamma_{\rm vdW}$ are the 
natural, Stark, and Van der Waals damping parameters, respectively. They are
given by the input parameters {\tt agam}, {\tt gs}, and {\tt gw}, respectively,
specified in the line list, {\tt fort.19}. They are given, after Kurucz (1970), by:
\begin{itemize}
\item Natural broadening:
\begin{equation}
\Gamma_{\rm rad} = \left\{ \begin{array}{ll}
10^{\tt agam},&  {\rm for}\ {\tt agam} > 0, \\ [1pt]
2.67\times 10^{-22}\nu_0^2,& {\rm for}\ {\tt agam} =0,
\end{array} \right.
\end{equation}
\item{Stark broadening}
\begin{equation}
\Gamma_{\rm Stark} = \left\{ \begin{array}{ll}
10^{\tt gs}\, n_{\rm e},&  {\rm for}\ {\tt gs} > 0, \\ [1pt]
10^{-8} n_{\rm eff}^{5/2}\, n_{\rm e},& {\rm for}\ {\tt gs} =0,
\end{array} \right.
\end{equation}
where $n_{\rm eff} \equiv Z_I [E_H/(E_{I}-E_{j)}]^{1/2}$ is the effective
quantum number of the upper level, $j$, of the transition in ion $I$, with the 
excitation energy $E_{j}$ and the ionization energy $E_I$,
$Z_I$ is the effective  charge ($Z_I=1$ for neutrals) of the ion $I$,  $E_H$ 
is the ionization energy of hydrogen, and $n_{\rm e}$ is the electron density.
\item{Van der Waals broadening}
\begin{equation}
\Gamma_{\rm vdW} = \left\{ \begin{array}{ll}
10^{\tt gw}\, c_w,&  {\rm for}\ {\tt gw} > 0, \\ [1pt]
4.5\times 10^{-9} (2.5\, n_{\rm eff}^4/Z^2)^{0.4}\,c_w,& {\rm for}\ {\tt gw} =0,
\end{array} \right.
\end{equation}
where $c_w=(N_{\rm H}+0.42 N_{\rm He}) (T/10^4)^{0.3}$.
\end{itemize}

\end{itemize}


\section*{Appendix B: Treatment of NLTE level populations in {\sc synspec}}
\addcontentsline{toc}{section}{Appendix B: Treatment of NLTE level populations 
in {\sc synspec}}

As explained above, for explicit species in a NLTE model atmosphere, 
the level populations used for evaluating the bound-free
are free-free transitions are taken from the model atmosphere generated
by {\sc tlusty}. 
On the other hand, the line list contains, for a given line, the $J$-values
and level energies for the lower and the upper level of the corresponding transition.
Based on this information, {\sc synspec} makes an association of the levels
being specified in the line list and the explicit levels that are specified through the
{\sc tlusty} input atomic data files. 

An association of levels specified in the line list, and the explicit levels specified
by the {\sc tlusty} standard input is done differently for light elements which do not
use the concept of superlevels and superlines, and for the iron-peak elements
which do use this concept.

\subsection*{Light elements}

For a given line, let $\ell$ denote a lower level referred to
in the line list, with its $J$-value and energy given as $J_\ell$ and $E_\ell$, 
respectively, and analogously for its upper level $u$.
The statistical weights and energies of the levels of the corresponding ion,
given by the {\sc tlusty} atomic data input, are $g_i$ and $\widetilde{E}_i$, 
$i=1,\ldots,N\!L$, respectively.
The task is to find which of the levels $i$ from the {\sc tlusty} input
correspond to $\ell$, this level is denoted as $I$.
The association $\ell \rightarrow I$ is done such as
$I=1$ for $E_\ell \leq \widetilde E_2/2$, and 
$(\widetilde E_I+\widetilde E_{I-1})/2 \leq E_\ell < (\widetilde E_I+\widetilde E_{I+1})/2$ otherwise.
The corresponding population of
level $\ell$ is then $n_\ell = n_I (2J_\ell+1)/g_I$, because $2J_\ell+1$ represents the
statistical weight of level $\ell$.  In other words, {\sc synspec} selects the
{\sc tlusty} explicit level which has the closest energy to the level specified by the line list,
and scales its population by the corresponding ratio of the statistical weights. 
Such scaling is necessary because
{\sc tlusty} often takes  for a ``level" a set of several actual energy eigenstates
of an atom. 

Finally, if $E_\ell > E_{N\!L}$, where $E_{N\!L}$ is the energy of the highest
level of the given ion considered by {\sc tlusty}, $n_\ell$ is set either to its LTE value
(if INLTE=1, where INLTE is specified in the file {\tt fort.55}); 
or $I=N\!L$. i.e., all such levels have the same $b$-factor as level $N\!L$ 
(if INLTE=2). 
A completely analogous procedure is done for level $u$.

Usually, the populations of the highest explicit levels are close to LTE anyway,
so these two options yield essentially the same results. In the (rare) case where the 
highest explicit level still departs significantly from LTE, the first option,
INLTE=1, may lead to wrong results that exhibit themselves by producing
spurious emissions. In such a case, the second option is preferable since 
it usually avoids these emissions. We stress that both options are just approximations;
the only exact way to deal with this problem is to improve model atoms by 
adding more high explicit energy levels. However, one may still face problems 
because necessary atomic data for such levels are usually poorly known.

The whole process is done automatically by {\sc synspec}, and
is thus completely transparent to the user.
In rare cases when the automatic association produces wrong results (for instance, 
if there are many different levels with very close energies but with different multiplicity),
one has to resort to the manual setting of explicit level indices as shown in \S\,\ref{linelist}.

\subsection*{Iron-peak elements}

Using 
he association of levels described above for the iron-peak elements may yield inaccurate 
results because a ``level' refereed to in the standard input represents a superlevel,
with a generally large span of actual energy levels. Importantly, the superlevels
are composed of levels with the same parity. When using an averaged energy
of a superlevel to associate it to the level energy specified in the line list, 
an information about a parity is not used.

Therefore, a better strategy is to provide an additional input file, called {\tt superlevels}, 
that contains an information about the energy limits of superlevels for both parities.
This information is also contained in the atomic data files for the individual
ions of the iron-peak elements treated explicitly (usually just Fe). This is explained
in detail in Paper~III, \S\,\refionsup. For now, the structure of the 
file is best explained 
on the following example, which provides a necessary input file when producing
synthetic spectra for the {\sc bstar2006} model atmospheres: We display here only the
beginning of the file {\tt superlevels}, the complete file is also a part of the
standard distribution of {\sc tlusty} \\ [2pt]

\hrule
\hrule
\begin{verbatim}
4         ! number of ions with superlevels
26 01     ! atomic number and charge of the first such ion (Fe II)
23        ! number of even-parity superlevels
  1000.   ! upper energy limit [in cm^-1] of the first superlevel
  5000.   ! analogously for higher superlevels
  1000.
  5000.
 10000.
 20000.
 25000.
 30000.
 35000.
 40000.
 45000.
 52000.
 65000.
 75000.
 82000.
 90000.
101000.
112000.
131000.
150000.
165000.
180000.
200000.
220000.
250000.
13         ! number of odd-parity superlevels
 40000.    ! energy limits, analogous as before
 50000.
 55000.
 70000.
 80000.
100000.
116000.
131000.
155000.
180000.
210000.
240000.
280000.
\end{verbatim}
\hrule
\hrule
\bigskip
This section contains data for Fe~II, the complete file {\tt superlines.ener} contains
analogous records for Fe~III, IV, and V.


\section*{Appendix C: Fine points of running {\sc synspec}}
\addcontentsline{toc}{section}{Appendix C: Fine points of running {\sc synspec}}

\label{fine}

A run of {\sc synspec} is composed of several basic steps:
\begin{itemize}
\item  
Initialization. The code reads the input model atmosphere and prepares
a number of depth-dependent, but frequency independent quantities, such 
as the solution of the equation of state (i.e., total populations of the individual 
ionization degrees for all selected chemical species), and a number of other
auxiliary quantities.
\item  
Reading of the line list(s) and a selection of lines that are taken into account.
Although one could in principle consider all lines that appear in the line list,
some sort of a selection of lines is useful because the line list contains
data for essentially all ionization stages of all the individual atoms, while only
some ionization stages are populated enough that their lines may 
contribute to the total opacity.

The selection of lines is done based on two input parameters appearing
in the file {\tt fort.55}, namely {\tt idstd} and {\tt relop}, and is done as follows:
Parameter {\tt idstd} specifies the index of the characteristic depth, as 
mentioned above in \S\,\ref{fort55}. The code computes a ratio of
the continuum opacity at the (approximate) position of the line and the 
line-center opacity, 
$r\equiv \kappa_{d}^{\rm line}(\lambda_0)/\kappa_{d}^{\rm cont}(\lambda_0)$.
If $r <$ {\tt relop}, the line is rejected. Here the subscript $d$, equal to {\tt idstd}, refers
to evaluating the opacities at depth {\tt idstd}. In some special cases, for instance
for atmospheres with a strong temperature rise toward the surface, the depth point
with index {\tt idstd} may have much lower temperature than that of the upper layers, 
and some lines which may contribute significantly to the opacity there may be rejected.
In that case the cure is to lower the parameter {\tt relop} to very low values, say
$10^{-20}$ to $10^{-30}$, in which case the code selects essentially all lines included
in the line list, many of them being completely negligible, but one can be sure that
important lines contributing in the upper layers are not rejected.
\item  
The code then proceeds sequentially through a number of analogous steps.
each computing a synthetic spectrum for a small interval of wavelengths, called
``set". There are several substeps:
\begin{itemize}
\item  
Setting the wavelength points: One goes through the list of selected lines,
from the lower wavelength to higher. For each selected line, the code places
a wavelength point in the line center, and in the midpoint between the current
line and the next one. The code then checks whether the distance of such points 
is larger than the maximum allowable distance of the points set through the
input parameter {\tt space} -- see \S\,\ref{fort55}. If so, the code sets additional 
wavelength points, equidistantly between the already established points, so that
their separation becomes smaller than {\tt space}.
In such a way it is guaranteed that no line center nor a window between two lines
is omitted, and one still achieves a desired wavelength resolution. 
\item  
Establishing a ``set":
Once the number of such wavelength points reaches some preset value,
taken by default to be 120, the setting of the wavelength points for this set
is completed. 
Subsequently, the depth-dependent continuum opacity is computed exactly
for the endpoints of the set, and is linearly interpolated for the inner
wavelength points.
This procedure is quite reasonable because it is faster than computing all the
wavelength-dependent continuum opacities exactly, and is sufficiently
accurate because the continuum opacity varies very
slowly with wavelength. Moreover, since the typical value of {\tt space} is around
0.01 \AA, the length of the set is about 1 \AA, the linear interpolation in
wavelengths is completely satisfactory.
\item  
Finally, computing the synthetic spectrum:
The code proceeds wavelength by wavelength, and for all depth points
it computes the total line opacity, adds the interpolated continuum opacity, and 
solves the radiative transfer equation.
\end{itemize}
\end{itemize}


\section*{Appendix D: Auxiliary {\sc synspec} output files}
\addcontentsline{toc}{section}{Appendix D: Auxiliary {\sc synspec} output files}

Here we describe in more detail two auxiliary output files from {\sc synspec}
that are interesting for some particular purposes. The first one is the line
identification table, {\tt fort.12}, and the second one is  the equivalent width
table, {\tt fort,16}. We shall describe them in turn.

\subsection*{Line identification table -- {\tt fort.12}}

The rationale for producing this table is a simple fact that one often needs
to know which actual lines contributed to the individual features seen in the 
synthetic spectrum. 

Let us take an example of a sample synthetic spectrum shown in Fig.2, namely
a short wavelength interval between 1400 and 1410 \AA, for an LTE model
atmosphere with $T_{\rm eff} = 35,000$ K, and $\log g=4$. We show the first
five records of the file {\tt fort.12} where we skip the first two columns which
contain some internal indices of lines, and which are kept for downward
compatibility without having any immediate practical meaning:
\begin{verbatim}
1400.160  Ni III  -1.27  156852.997   1.81E-04   0.0      0  0 51
1400.160  Co IV   -2.08  182501.396   7.57E-04   0.0      0  0 51
1400.195  Ni IV   -0.64  151574.707   3.08E+00  18.1  **  0  0 47
1400.237  Fe V     0.09  195196.298   1.01E+01  25.9  **  0  0 45
1400.237  Fe V    -1.36  219486.914   1.18E-01   1.6   *  0  0 45
\end{verbatim}
The individual columns represent:\\
-- wavelength [\AA]\\
-- indication of an atom and ion\\
-- $\log gf$\\
-- excitation energy of the lower level [cm${}^{-1}$]\\
-- ratio of the line-center opacity to the continuum opacity at the
standard depth\\
-- approximate equivalent width [m\AA]\\
-- an indication of the line strength (the more stars, the stronger the line)\\
-- the next two numbers are the indices of the lower and upper levels of the
transition for NLTE models. For LTE models, of for levels that are treated in LTE
in otherwise NLTE models, they are set to 0\\
-- the last number is the index of the depth where the monochromatic optical
depth in the line center is closest to $2/3$; i.e., roughly speaking, the index of the
depth of formation of the given line.
\smallskip

The approximate equivalent width,  $W$, is computed using classical expressions
(e.g. Hubeny \& Mihalas 2014, \S\,17.5), namely
\begin{equation}
W = \left\{ \begin{array}{ll}
(\sqrt\pi/2)\beta_0 [ 1 -(\beta_0/\sqrt{2}) + (\beta_0^2/\sqrt{3})] & \beta_0<1.2 \\ [4pt]
\sqrt{\ln\beta_0} & 1.2 \leq \beta_0 \leq 55 \\ [4pt]
(1/2) \sqrt{\pi a_0 \beta_0} & \beta_0 > 55.
\end{array} \right.
\end{equation}
where $\beta_0 = \chi_0/\kappa_c$ is the ratio of the line-center to the continuum
opacity, and $a_0$ is the Voigt damping parameter, both computed at the standard depth.

This information can be used to identify the spectral features in the predicted
spectrum. In fact, this is done in the above-mentioned IDL package {\sc synplot}.
As an illustration, we present in Fig. 4 an analogous plot to Fig.2, produced with
{\sc synplot}, that marks the individual predicted features, using the information from
{\tt fort.12}.
\begin{figure}[h]
\begin{center}
\label{fig2}
\includegraphics[width=4.7in]{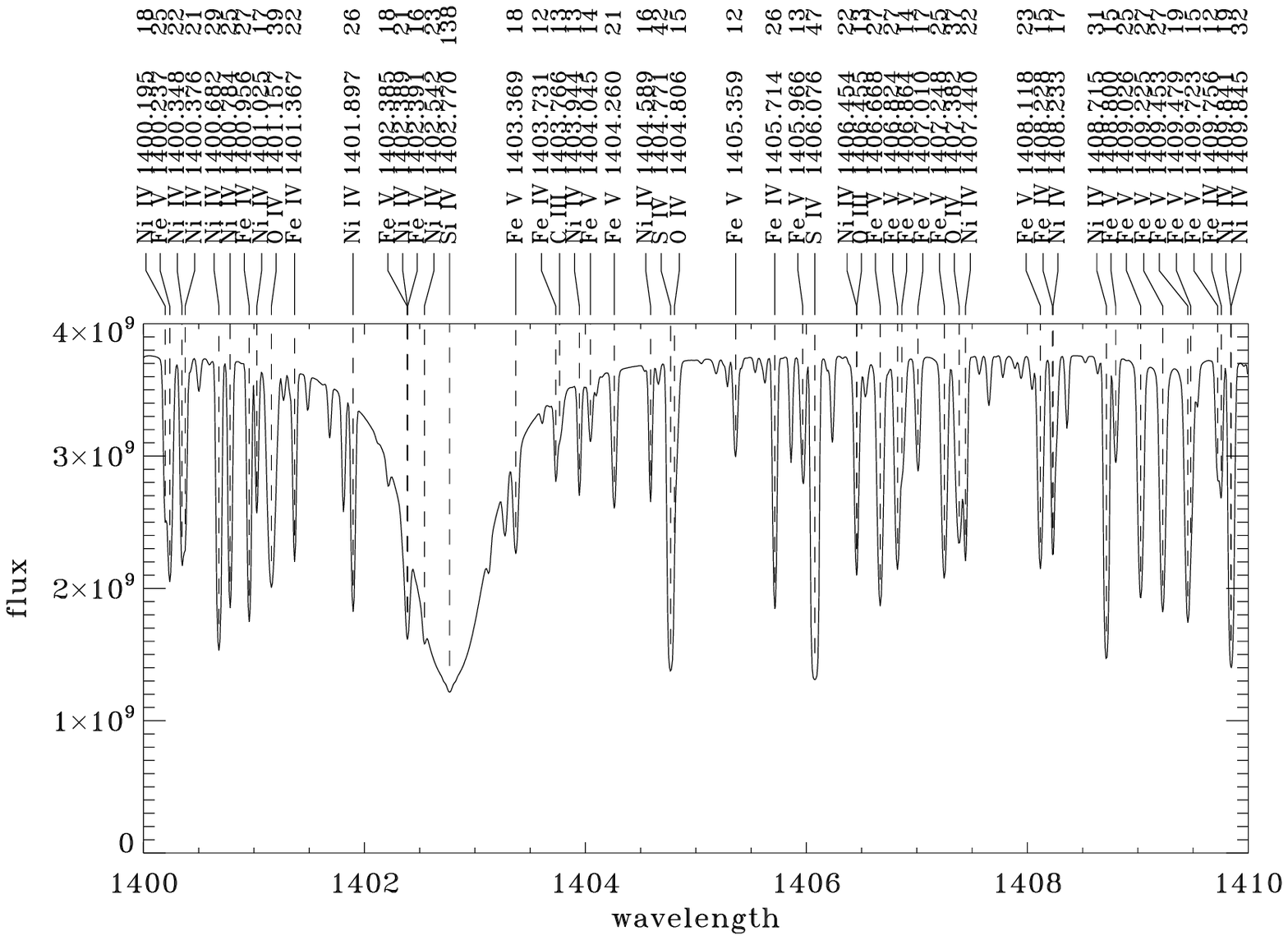}
\caption{Example of a synthetic spectrum for a simple LTE H-He model,
analogous to Fig. 2, with line annotations, produced by  {\sc synplot}.
The wavelength is in \AA,
and the flux is represented by the first moment of the specific intensity
$H_\lambda$ in erg~cm${}^{-2}$s${}^{-1}$\AA${}^{-1}$.
Labels at the top of the figure, (specification of the atom and ion, 
wavelength, and approximate equivalent width)  are taken directly from the file {\tt fort.12}.}
\end{center}
\vspace{-1em}
\end{figure}

\subsection*{Equivalent widths table -- {\tt fort.16}}

The classical concept of line equivalent width is still used in many spectroscopic
studies. However, this concept is only useful if one deals with isolated lines. In reality,
lines are very often blended, in particular in the UV spectral region. Consequently,
{\sc synspec} does not provide equivalent widths for individual lines, but rather
equivalent widths of the individual ``sets", i.e., the individual elementary wavelength
regions.

The run of {\sc synspec} used to produce Fig.\,2 has generated file {\tt fort.16}, which
looks like this:
\begin{verbatim}
    1400.000    1401.018     173.2    173.2      173.2     173.2
    1401.018    1402.017     167.3    167.3      340.6     340.6
    1402.017    1402.977     448.4    448.4      788.9     788.9
    1402.977    1403.928     239.6    239.6     1028.5    1028.5
    1403.928    1404.935     129.1    129.1     1157.7    1157.7
    1404.935    1405.918      60.8     60.8     1218.4    1218.4
    1405.918    1406.972     180.0    180.0     1398.4    1398.4
    1406.972    1407.942      98.4     98.4     1496.8    1496.8
    1407.942    1408.904     106.2    106.2     1603.0    1603.0
    1408.904    1409.846     160.5    160.5     1763.5    1763.5
    1409.846    1410.000      21.9     21.9     1785.5    1785.5
\end{verbatim}
The individual columns have the following meaning:\\
-- initial wavelength of the interval $\lambda_0$ [\AA]\\
-- ending wavelength of the interval $\lambda_1$[\AA]\\
-- equivalent width [m\AA] of this interval (EQW)\\
-- a modified equivalent width, defined such thaf any emission features, which
would produce a negative contribution to the equivalent width, are cut off (MEQW).
Obviously, for a synthetic spectrum with only absorption lines, EQW and MEQW
are identical.\\
-- cumulative equivalent width (a sum over all previous intervals)\\
-- cumulative modified equivalent width.\\

\noindent Specifically,
\begin{equation}
{\rm EQW} = \int_{\lambda_0}^{\lambda_1} 
[1-(F_\lambda/F_\lambda^{\rm cont}) ]\,
d\lambda,\nonumber
\end{equation}
and
\begin{equation}
{\rm MEQW} = \int_{\lambda_0}^{\lambda_1} 
[1-\min(F_\lambda/F_\lambda^{\rm cont}, 1) ]\,
d\lambda.\nonumber
\end{equation}
%


\section*{Appendix E: Program {\sc rotin}}
\addcontentsline{toc}{section}{Appendix E: Program {\sc rotin}}

This is a very simple program that performs the rotational and instrumental
convolution of the original stellar spectrum. The instrumental profile is
assumed to be Gaussian.

The general structure of the standard input file is, using the variable names exactly
as they are in the {\sc rotin} source code
\begin{verbatim}
fname7   fname17  fnout
vrot     chard    stepr
fwhm     stepi
alam0    alam1    irel
\end{verbatim}
Here is a brief explanation of the input parameters:\\ [2pt]
$\bullet$ {\tt fname7} -- name of the file containing the detailed spectrum 
(output {\tt fort.7} from {\sc synspec}).\\ [2pt]
$\bullet$ {\tt fname17} -- name of the file containing the continuum flux 
(output {\tt fort.17} from {\sc synspec}).\\ [2pt]
$\bullet$ {\tt fnout} -- filename of the output convolved spectrum.\\ [2pt]
$\bullet$ {\tt vrot} -- rotational velocity $v \sin i$ in km/s.\\ [2pt]
$\bullet$ {\tt chard} -- characteristic distance between two neighboring wavelength 
points  of the original spectrum - in \AA.\\
-- if $=0$, program sets up the default value of  0.01 \AA.\\ [2pt]
$\bullet$ {\tt stepr} -- wavelength step for evaluation rotational convolution;\\
-- if $=0$, the program sets up default (the wavelength
interval corresponding to the rotational velocity divided by 3.)\\
-- if $<0$, convolved spectrum calculated on the original
(detailed) {\sc synspec} wavelength mesh.\\ [2pt]
$\bullet$ {\tt fwhm} -- full width at half maximum for a Gaussian instrumental profile
(in \AA).\\ [2pt]
$\bullet$ {\tt stepi} -- wavelength step for evaluating instrumental convolution (in \AA)\\
- if $=0$, the program sets up the default value of {\tt fwhm}/10.\\
- if $<0$, convolved spectrum calculated on the previous wavelength mesh:
used in rotational convolution.\\ [2pt]
$\bullet$ {\tt alamo, alam1} -- starting and ending wavelength for the convolved
spectrum (in \AA).\\ [2pt]
$\bullet$ {\tt irel} -- a switch for setting an evaluation of the absolute or relative 
spectrum:\\
-- if $=0$, an output is the absolute convolved spectrum, in the same units as 
the original stellar spectrum on {\tt fname7} \\
-- if $=1$, an output is the spectrum normalized to the continuum (from input  
continuum flux on {\tt fname17}).

\medskip
The output file, with the name specified by the parameter {\tt fnout}, is a simple
table of wavelength (in \AA) versus flux, either absolute or normalized to continuum, 
as specified by the input -- see above.

\medskip
For completeness, the convolutions for the rotational and instrumental broadening
are evaluated as 
\begin{equation}
F^{\rm conv}(\lambda) = \int_{\lambda-\Delta\lambda_0}^{\lambda+\Delta\lambda_0}
G(\lambda- \lambda^\prime) F(\lambda^\prime)\, d\lambda^\prime,
\end{equation}
where $F$ is the radiation flux, and $G$ the corresponding kernel function.
For the rotational broadening, $G$ is given by (e.g., Gray 2008)
\begin{equation}
G(\Delta\lambda) = \frac{2(1-\epsilon)
\left[1-\left(\Delta\lambda/\Delta\lambda_0\right)^2\right]^{1/2} +
(\pi\epsilon/2)\left[1-\left(\Delta\lambda/\Delta\lambda_0\right)^2 \right] }
{\pi \Delta\lambda_0(1-\epsilon/3)},
\end{equation}
where $\lambda_0=(\lambda\,v \sin i)/c$, $\epsilon$ is the limb-darkening coefficient,
which is hardwired  in {\sc rotin} as $\epsilon=0.6$. It would be possible to consider
it as a free parameter, but we decided not to. The reason is that for evaluating an 
accurate rotationally broadened spectrum it is preferable to perform an integration over 
the stellar surface exactly, using angle-dependent specific intensities of radiation, 
which can also be produced by {\sc synspec}, instead of flux.

For the instrumental convolution, the kernel is given by
\begin{equation}
G(\Delta\lambda) = c_1 \exp[-{\Delta\lambda/\Delta\lambda_I}^2],
\end{equation}
where $\Delta\lambda_I = {\rm FWHM}/(2\sqrt{\ln 2}) = 0.60056 \times {\rm FWHM}$, 
and $c_1=1/(\sqrt{\pi} \Delta\lambda_I)$. The integration limit is formally 
$\Delta\lambda_0=\infty$; for a numerical integration it is chosen 
$\Delta\lambda_0=3\, \Delta\lambda_I$.

\section*{References}
\addcontentsline{toc}{section}{References} 

\def\reference{\par \leftskip20pt \parindent-20pt\parskip4pt}
\noindent
\reference Allard,  N., \& Koester, D. 1992, A\&A, 258, 464.
\reference Auer, L.H., \& Mihalas, D, 1969, ApJ, 158, 641.
\reference Auer, L.H., \& Mihalas, D, 1970, ApJ, 160, 233.
\reference Auer, L.H., \& Mihalas, D, 1972, ApJS, 24, 193.
\reference Barnard, A., Cooper, J., Smith, E. 1974, JQSRT, 14, 1025.
\reference Cunha, K., Hubeny, I., \& Lanz, T. 2006, ApJ, 647, L143.
\reference Dimitrijevic, M., \& Sahal-Br\'echot, S. 1984, JQSRT, 31, 301.
\reference Freudenstein, \& Cooper, J. 1978, ApJ, 224, 1079.
\reference Gray, D.F., 2008, {\it The Observation and Analysis of Stellar Photospheres},
Cambridge Univ. Press, Cambridge, 3rd edition.
\reference Hubeny, I. 1988, Computer Physics Commun. 52, 103.
\reference Hubeny, I., Hummer, D.G., \& Lanz, T. 1994, A\&A, 282, 151.
\reference Hubeny, I., \& Lanz, T. 1995, ApJ, 439, 875.
\reference Hubeny, I., \& Lanz, T. 2011, SYNSPEC, Astrophys. Source Code Library,
1189.822.
\reference Hubeny, I., \& Lanz, T. 2017b,  {\sc tlusty} User's Guide II: Reference Manual,
(Paper~II)
\reference Hubeny, I., \& Lanz, T. 2017c,  {\sc tlusty} User's Guide III: Operational Manual,
(Paper~III)
\reference Hubeny, I., Lanz, T., \& Jeffery, C.S. 1994, in Newsletter on
   Analysis of Astronomical Spectra No. 20, ed. C.S. Jeffery,
   St. Andrews Univ., p.30.
\reference Hubeny, I. \& Mihalas, D. 2014, {\it Theory of Stellar Atmospheres},
Princeton Univ. Press, Princeton.
\reference Kurucz, R.L. 1979, ApJS, 40, 1.
\reference Mihalas, D., \& Auer, L.H., 1970, ApJ, 160, 1161.
\reference Mihalas, D., Heasley, J., \& Auer, L.H., 1975, Technical Report 
NCAR-TH/STR-104, Boulder, Colorado.
\reference Lanz, T., Cunha, K., Holtzman, J., \& Hubeny, I. 2008, ApJ, 678, 1342.
\reference Lanz, T., \& Hubeny, I. 2003, ApJS, 146, 417.
\reference Lanz, T., \& Hubeny, I. 2007, ApJS, 169, 83.
\reference Lemke, M., 1997, A\&AS, 122, 285.
\reference Schoening, T., \& Butler, K. 1989, A\&AS, 78, 51.
\reference Shamey, L.1969, PhD thesis, University of Colorado.
\reference Tremblay, P.-E., \& Bergeron, P. 2009, ApJ, 696,1755.

\newpage

\def\noreference{\par \leftskip0pt \parindent0pt\parskip4pt}

\noreference

\end{document}